\documentclass[a4paper]{article}
\usepackage[utf8]{inputenc}
\usepackage{amsmath}
\usepackage{amsfonts}
\usepackage{amssymb}
\usepackage{graphicx}
\author{Etienne Mangaud, Benjamin Rotenberg \\
Sorbonne Universit\'e\\  CNRS \\ Physicochimie des \'electrolytes et Nanosyst\`emes Interfaciaux \\
 F-75005 Paris, France}
\title{Sampling mobility profiles of confined fluids with equilibrium molecular dynamics simulations}

\begin{document}

\maketitle


\section*{Abstract}
We show how to evaluate mobility profiles, characterizing the transport of confined fluids under a perturbation, from equilibrium molecular simulations. The correlation functions derived with the Green-Kubo formalism are difficult to sample accurately and we consider two complementary strategies: improving the spatial sampling thanks to a new estimator of the local fluxes involving the forces acting on the particles in addition to their positions and velocities, and improving temporal sampling thanks to the Einstein-Helfand approach instead of the Green-Kubo one. We illustrate this method on the case of a binary mixture confined between parallel walls, under a pressure or chemical potential gradient. All equilibrium methods are compared to standard non-equilibrium molecular dynamics (NEMD) and provide the correct mobility profiles. We recover quantitatively fluid viscosity and diffusio-osmotic mobility in the bulk part of the pore. Interestingly, the matrix of mobility profiles for local fluxes is not symmetric, unlike the Onsager matrix for the total fluxes. Even the most computationally efficient equilibrium method (Einstein-Helfand combined with the force-based estimator) remains less efficient than NEMD to determine a specific mobility profile. However, the equilibrium approach provides all responses to all perturbations simultaneously, whereas NEMD requires the simulation of several types of perturbations to determine the various responses, each with different magnitudes to check the validity of the linear regime. While NEMD seems more competitive for the present example, the balance should be different for more complex systems, in particular for electrolyte solutions for the responses to pressure, salt concentration and electric potential gradients.

\section*{Introduction}

Transport in nanochannels has been an ever-growing topic of interest in the past decades\cite{eijkelNanofluidicsWhatIt2005}. Technological advances have enabled the design and setup of smaller devices, and emphasized the role of the channel surface and its interplay with the fluid on the transport properties\cite{bocquetNanofluidicsBulkInterfaces2010,schochTransportPhenomenaNanofluidics2008,sparreboomTransportNanofluidicSystems2010}. Importantly, surface-driven flows can be generated by an electric potential, concentration or temperature gradient, which correspond to electro- \cite{rotenbergElectrokineticsInsightsSimulation2013}, diffusio- \cite{marbachOsmoticDiffusioosmoticFlow2017,leeNanoscaleDynamicsSurface2017} and thermo-osmosis \cite{fuWhatControlsThermoosmosis2017}, respectively.
As an example, diffusio-osmosis, which originates from a force imbalance at the interface\cite{leeOsmoticFlowFully2014}, could be exploited for many applications\cite{marbachOsmosisMolecularInsights2019}, notably harvesting blue energy\cite{siriaGiantOsmoticEnergy2013}. 

The potential of these phenomena to manipulate fluids on the nanoscale has stimulated the development of theoretical and numerical tools to clarify how the interactions between the fluid and the surface on the molecular scale result in the observed mass or charge transport on larger scales.
While the validity of continuum hydrodynamics has proven to hold down to remarkably small scales (a few molecular diameters\cite{bocquetNanofluidicsBulkInterfaces2010}), surface effects modify the boundary conditions, \emph{e.g.} via a slip velocity. A microscopic approach is necessary to investigate and predict the local response of the interfacial fluid to various perturbations, which in turn controls the flow far from the interface. 

Here we consider specifically the linear response of a confined binary mixture (solute $A$ in solvent $B$) to a small pressure gradient $-\nabla P$ and/or chemical potential gradient $-\nabla \mu$, but the present work can be straightforwardly generalized to any kind of (small) perturbation, in particular to the response of electrolytes under pressure, salt concentration and electric potential gradients.
For sufficiently small perturbations, the system responds linearly and the various fluxes are written in Onsager's framework for the thermodynamics of irreversible processes by introducing a symmetric matrix $\mathcal{L}$ of transport coefficients\cite{marbachOsmosisMolecularInsights2019}:
\begin{equation}
\left(\begin{array}{c}  Q \\  J_A - c_A^*  Q  \end{array} \right) = \mathcal{L} \left(\begin{array}{c} -\nabla P \\ -\nabla \mu  \end{array} \right)
\label{eq:onsager}
\end{equation}
where $Q$ is the volume flux, $J_A$ the solute flux, and $ J_A - c_A^*  Q$ the excess flux of solute, with $c_A^*$ a reference solute concentration discussed in more detail below. The elements of the Onsager matrix of transport coefficients $\mathcal{L}$:
\begin{equation}
\mathcal{L} = \left(\begin{array}{cc}  \mathcal{L}_{11} & \mathcal{L}_{12}\\ \mathcal{L}_{21} & \mathcal{L}_{22} \end{array} \right) 
\label{eq:tr_mat}
\end{equation}
quantify the permeability ($\mathcal{L}_{11}$), diffusio-osmosis ($\mathcal{L}_{12}$), excess solute flux under pressure ($\mathcal{L}_{21}$) and Maxwell-Stefan diffusion ($\mathcal{L}_{22}$), respectively and Onsager's time-reversal symmetry relationship implies $\mathcal{L}_{12}=\mathcal{L}_{21}$. 

While the average fluxes are sufficient to characterize the macroscopic response of the system, they do not underline the importance of the local response of the interface on the latter. Indeed, depending on the nature of the perturbation the balance between the various fluxes differ and subtle differences in the interaction of all components of the fluid with the walls result in qualitatively different flow profiles. In the case of a fluid confined between two parallel walls (slit pore), to which we will restrict our discussion in the present work, a pressure gradient along the walls results in the well-known parabolic, Hagen-Poiseuille velocity profile, while the diffusio-osmotic flow induced by a chemical potential gradient along the walls, when the components of the fluid have different affinity for the walls, is plug-like (\textit{i.e.} uniform beyond the interfacial region) and the bulk pressure remains constant. In turn, the distribution of the species through the pore combined with the different shapes of the flow profiles yield different solute and solvent fluxes for each type of perturbation. 

Therefore, in order to understand the role of molecular interactions at the interface, it is relevant to also consider the local fluxes induced by macroscopic perturbations, defined in this slit geometry as a function of their position $z$ in the direction perpendicular to the walls. In the linear response regime, the local responses are governed by the following matrix of local transport coefficients, or mobilities, $\mathcal{M}_{ij}(z)$, defined by
\begin{equation}
\left(\begin{array}{c} q(z) \\ j_A(z) - c_A^* q(z)  \end{array} \right) = \mathcal{M} (z)\left(\begin{array}{c} -\nabla P \\ -\nabla \mu  \end{array} \right)
\label{eq:mobility}
\end{equation}
where $q$ and $j_A$ are the local counterparts of the total fluxes $Q$ and $J_A$ in Eq.~\ref{eq:onsager}. These mobilities fully characterize the flow and excess solute flux profiles in response to the various perturbations. Unlike $\mathcal{L}$, this local mobility matrix is not, in general, symmetric -- \emph{i.e.} $\mathcal{M}_{21}(z) \neq \mathcal{M}_{12}(z)$. However the macroscopic transport coefficients in Eq.~\ref{eq:tr_mat} are obtained as the average of the local mobilities
\begin{equation}
\mathcal{L}_{ij} =  \frac{1}{H}\int_0^H {\rm d}z\, \mathcal{M}_{ij} (z) 
\; ,
\label{eq:LijfromMij}
\end{equation}
with $H$ the width of the slit pore (distance between the walls confining the fluid), and should obey the symmetry relation.

Molecular dynamics (MD) simulations have now become a standard tool to investigate the properties of matter, including transport, on the microscopic scale. In non-equilibrium molecular dynamics (NEMD) simulations, the system is submitted to an external perturbation and the local solute and solvent fluxes are sampled at steady-state. This approach has long been exploited to predict the response of confined fluids to an applied pressure gradient or an electric field, by including an external force or electric field in the equations of motion.\cite{karniadakisMicroflowsNanoflowsFundamentals2005,travisDepartureNavierStokesHydrodynamics1997,botanHydrodynamicsClayNanopores2011} It was also more recently extended to the case of diffusio-osmotic flows using forces related to the chemical nature of the various species\cite{yoshida_generic_2014,yoshidaOsmoticDiffusioosmoticFlow2017}, which avoid some difficulties of other mechanical approaches to mimic the effect of concentration gradients\cite{liuPressureGradientsFail2018}. 

In practice, one applies one type of perturbation and samples the fluxes of all species as a function of the position with respect to the surfaces; this is then repeated for each type of perturbation.\cite{bocquet_hydrodynamic_1994,jolyLiquidFrictionCharged2006,botan_hydrodynamics_2011,ameur_slippage_2013,botan_how_2013,rotenbergElectrokineticsInsightsSimulation2013,yoshida_generic_2014,ganti_molecular_2017,simonnin_mineral_2018,fu_what_2017,fu_giant_2019} While natural and efficient, the NEMD approach brings the system out of equilibrium and requires appropriate thermostatting strategies, on which the resulting fluxes should not depend.\cite{yong_thermostats_2013,sam_water_2017,ruiz-franco_effect_2018} In addition, the mobilities are only defined in the linear response regime, and several simulations need to be performed to find the good compromise between the validity of the linear response regime (implying small perturbation) and signal-to-noise ratio (larger for large perturbation).

Equilibrium molecular dynamics (EMD) simulations provide in principle an attractive alternative to predict the transport properties. Using linear response theory, it is indeed possible to derive Green-Kubo expressions for all the responses to various types of perturbations, in which transport coefficients are expressed as time integrals of equilibrium correlation functions. All the latter can therefore be computed simultaneously from the same equilibrium trajectories. In addition, the system evolves without the need to dissipate the power introduced by the external perturbation. This approach has already been used to determine the components of the Onsager matrix $\mathcal{L}$ (see Eq.~\ref{eq:onsager}) for fluid confined in slit pores or as a liquid film on a substrate.\cite{yoshidaMolecularDynamicsSimulation2014,yoshidaOsmoticDiffusioosmoticFlow2017} Its application to determine their local counterparts, \emph{i.e.} the components local mobility matrix $\mathcal{M}(z)$, remains however very challenging because it requires computing time correlation functions over long timescales, which are evaluated for a large number of bins corresponding to each position $z$. As a result, only a handful of studies have adopted this approach to determine the electro-osmotic flow profile induced by an electric field\cite{marryEquilibriumElectrokineticPhenomena2003,dufreche_molecular_2005}, or the Poiseuille flow induced by a pressure gradient\cite{agnihotriDisplacementsMeanSquaredDisplacements2014}.

Inspired by the recent development of improved estimators for the computation of local properties such as number, charge or polarization densities\cite{borgisComputationPairDistribution2013,schultzReformulationEnsembleAverages2016,delasherasBetterCountingDensity2018,colesComputingThreedimensionalDensities2019,schultzAlternativesConventionalEnsemble2019}, we explore in this work the possibility to compute the local mobility matrix $\mathcal{M}(z)$ from equilibrium MD simulations. We develop the theoretical framework to compute the profiles for all transport coefficients, in the case of a binary mixture confined in a slit pore, using the Green-Kubo formalism. The resulting expressions involve the integral of cross-correlation functions between local and global fluxes. We then introduce two improvements to compute them efficiently: firstly, we introduce a new estimator of these cross-correlation functions, which makes use of the instantaneous force acting on the atoms in addition to their position and velocity; secondly, we avoid the sampling of the time-correlation functions by deriving the ``Einstein'' counterparts of the Green-Kubo expressions, \emph{i.e.} using the displacements instead of the velocities, as proposed for the case of Poiseuille flow in Ref.~\cite{agnihotriDisplacementsMeanSquaredDisplacements2014}. The results obtained by this equilibrium route are compared to NEMD simulations, which we use as a reference to validate the proposed method.

The theoretical basis to compute the mobility matrix from equilibrium MD simulations is presented in Section~\ref{sec:theory}. Section~\ref{sec:simulations} then introduces the simulation details, with an emphasis on the case of the diffusio-osmotic response, which is the most difficult element of the mobility matrix to evaluate. The strategies to improve the sampling of the cross-correlation functions between local and global fluxes are introduced in Section~\ref{sec:improving} and the results presented in Section~\ref{sec:results}.

\section{Transport coefficients profiles from equilibrium simulations}
\label{sec:theory}

\subsection{Fluxes}
\label{sec:theoryfluxes}

We consider a binary mixture (solute $A$ in  solvent $B$) confined between two planar solid surfaces separated by a distance $H$ (see Fig.~\ref{fig:system}a). The response of the fluid is investigated for two kinds of perturbations, namely a pressure gradient or a chemical potential gradient, both applied in the direction $x$ parallel to the surface. Each of these perturbations induces a flow parallel to the plane but with different profiles in the direction $z$ normal to the surface, typically parabolic and plug-like velocity profiles for pressure-driven and diffusio-osmotic flow, respectively. These responses are intimately related to the equilibrium local structure and composition of the fluid inside the pore. The latter are characterized by the solute, solvent and total densities $\rho_A$, $\rho_B$ and $\rho$, defined as  
\begin{align}
\label{eq:denstot}
\rho(z) &= \rho_A(z) + \rho_B(z) \\
\rho_\nu(z) &= \frac{1}{S}\left\langle \sum_{i=1}^{N_\nu} \delta(z_i-z) \right\rangle
\label{eq:dens}
\end{align}
where $S$ is the surface area, $N_\nu$ is the number of particles of type $\nu = A, B$, $\delta$ is the Dirac delta function, $z_i$ is the position of particle $i$ in the direction perpendicular to the surfaces, and the brackets denote averages in the canonical ensemble (fixed volume $V$, system composition and temperature $T$). 

The microscopic observables corresponding to the local fluxes in Eq.~\ref{eq:mobility} can be defined from the instantaneous positions and velocities of the fluid particles: the local volume flux involves all the $N=N_A+N_B$ particles,
\begin{align}
q (z,t) &=  \frac{H}{N} \sum \limits_{i=1}^{N} v_{x,i}(t) \delta (z_i(t)-z) 
\; ,
\label{eq:q_z} 
\end{align}
with $v_{x,i}$ the $x$-component of the velocity of particle $i$, while the local solute particle fluxes are defined for each species as
\begin{align}
j_\nu (z,t) &=  \frac{1}{S} \sum \limits_{i=1}^{N_\nu} v_{x,i}(t) \delta (z_i(t)-z)
\; .
\label{eq:j_z} 
\end{align}
The volume flux is related to the total particle flux, $j(z,t)=j_A(z,t) + j_B(z,t)$, as $q(z,t)=\dfrac{V}{N}j(z,t)$.
The fluxes $Q$ and $J_A$ entering in Eq.~\ref{eq:onsager} are obtained from the local ones as:
\begin{equation}
  Q(t)  = \frac{1}{H} \int_0^H {\rm d}z\, q(z,t) = \frac{1}{N} \sum \limits_{i=1}^{N} v_{x,i}(t) 
  \label{eq:Qoft}
\end{equation}
and
\begin{equation}
 J_\nu(t) = \frac{1}{H} \int_0^H {\rm d}z\, j_\nu(z,t) = \frac{1}{V} \sum \limits_{i=1}^{N_\nu} v_{x,i}(t)
   \label{eq:Jnuoft}
\end{equation}
Finally, the reference concentration $c_A^*$ used in the definition of the excess solute flux (see Eq.~\ref{eq:onsager}) is defined as\cite{yoshidaOsmoticDiffusioosmoticFlow2017}
\begin{align}
\label{eq:cA}
c_{A}^*=\frac{N_A^b}{N_A^b+N_B^b} \frac{N}{V} =\frac{\alpha}{1+\alpha}\frac{N}{V}
\; ,
\end{align}
with $\alpha=\rho_A^b/\rho_B^b=N_A^b/N_B^b$, where $\rho_A^b$ and $\rho_B^b$ are the solute and solvent densities in the bulk region, and $N_A^b$ and $N_B^b$ the corresponding numbers of particles (see Section~\ref{sec:dof} for more details).

The transport coefficients defined by Eqs.~\ref{eq:onsager} and~\ref{eq:mobility} can be computed from the steady-state averages of the above instantaneous fluxes in NEMD simulations. Such an approach, already used in the references cited in the Introduction, provides a reference methodology. However, it requires several simulations for various strengths of each perturbation, in order to check the linear response of the system. In the next subsection, we introduce the mechanical perturbations associated with pressure and chemical potential gradients, as well as the corresponding Green-Kubo expressions of the transport coefficients for the linear response to these perturbations. These expressions feature correlation functions, computed from equilibrium MD simulations, \emph{i.e.} in the absence of mechanical perturbation. This method not only determines all transport coefficients simultaneously, it also allows it without the need to perform several non-equilibrium simulations to verify the validity of the response regime.

\subsection{Pressure gradient: Permeability and excess flux under pressure}
\label{sec:theorygradP}

A uniform and constant pressure gradient $\nabla P$ in the direction $x$ parallel to the walls can be simulated by applying a force ${\bf f}_i^P=-\frac{1}{\rho_0}\nabla P$, with $\rho_0=N/V$ the fluid density, on all particles. This perturbation corresponds to the Hamiltonian
\begin{equation}
\label{eq:hamiltonianP}
H_{P} = -\sum_{i=1}^{N} f_{x,i}^P x_i
= \left(\frac{V}{N}\nabla_x P\right) \sum_{i=1}^{N} x_i 
\; ,
\end{equation}
with $x_i$ the $x$ coordinate of particle $i$.
Following the standard derivations of linear response theory\cite{hansenTheorySimpleLiquids2006}, one  obtains the following Green-Kubo expression for the steady-state local volume flux:
\begin{align}
\left\langle q(z) \right\rangle &=  -\frac{1}{k_BT} \int_{0}^{+\infty} {\rm d}t\, \left\langle q(z,t) \dot{H}_{P}(0) \right\rangle
\nonumber \\
& = ({ -\nabla_x P}) \times \frac{V}{k_BT} \int_{0}^{+\infty} {\rm d}t\, \left\langle q(z,t) Q(0) \right\rangle
\; ,
\label{eq:qresponsetogradP}
\end{align}
where we used Eq.~\ref{eq:hamiltonianP} and the definition~\ref{eq:Qoft} to express $\dot{H}_{P}(0)$.
This allows the identification of the corresponding mobility entering in Eq.~\ref{eq:mobility} as
\begin{align}
\mathcal{M}_{11}^{GK}(z)&= \frac{  V }{ k_B T } \int_{0}^{+\infty} {\rm d}t\, {C_{11}}(t,z)
  \label{eq:m11}
\end{align}
with
\begin{equation}
C_{11}(t,z) = {C_{q Q}}(t,z) = \left\langle q(z,t)Q(0)\right\rangle
 \label{eq:c11}
\end{equation}
the cross-correlation between the local and average volume fluxes. Similarly, under the same perturbation, the mobility profile for the excess solute flux can be expressed as
\begin{align}
\mathcal{M}_{21}^{GK}(z) = \frac{  V }{ k_B T} \int_{0}^{+\infty} {\rm d}t\, {C_{21}}(t,z)
  \label{eq:m21}
\end{align}
with 
\begin{equation}
 {C_{21}}(t,z) = C_{j_A Q}(t,z) - c_A^* C_{q Q}(t,z)
 \; ,
 \label{eq:c21}
\end{equation}
where $C_{j_A Q}(t,z)$ is the cross-correlation between the local solute flux $j_A$ and average volume flux $Q$, and $c_A^*$ is defined by Eq.~\ref{eq:cA}.

\subsection{Chemical potential gradient: Diffusio-osmotic flow and Maxwell-Stefan diffusion}
\label{sec:dof}

 The mechanical description of the effect of a chemical potential gradient is not as straightforward as that of a pressure gradient. It is nevertheless possible and we follow here the approach of Ref.~\cite{yoshidaOsmoticDiffusioosmoticFlow2017} where a constant force ${\bf f}^\mu$ is applied to each solute particle and a force $-\alpha{\bf f}^\mu$ to each solvent particle, with $\alpha$ such that there is no net force on the fluid in the bulk region far from the walls, where the fluid is homogeneous. This implies $\alpha=\rho_A^b/\rho_B^b=N_A^b/N_B^b$, as introduced below Eq.~\ref{eq:cA}. This perturbation corresponds to a Hamiltonian
\begin{equation}
H_{\mu} = -\sum_{i=1}^{N} f_{x,i}^\mu x_i =
-f_x^\mu \left[ \sum_{i=1}^{N_A} x_i - \alpha  \sum_{i=1}^{N_B} x_i  \right] 
\label{eq:hamiltonianMu}
\end{equation}
As shown in Ref.~\cite{yoshidaOsmoticDiffusioosmoticFlow2017}, the effect of these applied forces corresponds to a chemical potential gradient 
$-\nabla_x \mu = \frac{N_A^b+N_B^b}{N_B^b} f_x^\mu = (1+\alpha) f_x^\mu$.
Therefore, linear response theory provides the analog of Eq.~\ref{eq:qresponsetogradP} for the steady-state local volume flux as
\begin{align}
\left\langle q(z) \right\rangle &=  -\frac{1}{k_BT} \int_{0}^{+\infty} {\rm d}t\, \left\langle q(z,t) \dot{H}_{\mu}(0) \right\rangle
\nonumber \\
& = \frac{{ -\nabla_x \mu}}{1+\alpha} \times \frac{V}{k_BT} \int_{0}^{+\infty} {\rm d}t\, \left\langle q(z,t) \left[J_A(0)-\alpha J_B(0)\right] \right\rangle
\; .
\label{eq:qresponsetogradMu}
\end{align}
Using $J_B=\frac{N}{V}Q-J_A$, this result can be rewritten to express the diffusio-osmotic mobility
\begin{align}
\mathcal{M}_{12}^{GK}(z) =  \frac{ V  }{ k_B T } \int_{0}^{+\infty} {\rm d}t\, C_{12}(t,z) 
\label{eq:m12}
\end{align}
with
\begin{align}
C_{12}(t,z) = {C_{q J_A}}(t,z)    -  c_{A}^* {C_{q Q}}(t,z) 
\; ,
\label{eq:c12}
\end{align}
where $C_{q J_A}$ is the cross-correlation between the local volume flux $q$ and the average solute flux $J_A$, and $c_A^*$ is defined by Eq.~\ref{eq:cA}.
Similarly, the same derivation for the excess solute flux $j_A(z) - c_A^* q(z)$ with the perturbating Hamlitonian $H_\mu$ results in the mobility:
\begin{align}
\mathcal{M}_{22}^{GK}(z) =  \frac{ V  }{ k_B T } \int_{0}^{+\infty} {\rm d}t\, C_{22}(t,z)
\label{eq:m22}
\end{align}
with
\begin{align}
C_{22}(t,z) = C_{j_A J_A}   -  c_{A}^* (C_{j_A Q} + C_{q J_A}) + (c_A^*)^2 {C_{q  Q}}
\label{eq:c22}
\end{align}
with the cross-correlations between local and average fluxes are defined as above. This element of the mobility matrix characterizes the relative transport of solute and solvent in the presence of a  chemical potential gradient and corresponds, up to differences in the definition of the transport coefficients, to the so-called Maxwell-Stefan diffusion.

Far from the walls, a chemical potential gradient results in opposite fluxes of solute and solvent but in no local net force on the bulk region of the fluid. The above mechanical description clarifies the role of the differential affinity of the solute and solvent for the walls in inducing the diffusio-osmotic flow: it is the different composition of the interfacial region (compared to the bulk), which result in a local net force accelerating the fluid; viscous momentum diffusion away from the interface then results in a non-zero fluid velocity even far from the interface. The role of the force balance also underlines the importance of finding a suitable ``bulk'' region in the confined system to define the corresponding bulk densities and ratio $\alpha$ (and corresponding reference concentration $c_A^*$) to predict the responses to a chemical potential gradient. This choice will be discussed in more detail in Section~\ref{sec:simulations}.
Finally, we note that all the mobility profiles, \emph{i.e.} the two types of responses to the two types of perturbations, can be computed from the same equilibrium simulations, by sampling the various correlation functions between local and average fluxes.


\section{How to improve the sampling of transport coefficients ?}
\label{sec:improving}

The correlation functions ${{C_{kl}}(t,z)}$ 
provide in principle a direct route to the elements of the local mobility matrix $\mathcal{M}(z)$ in Eq.~\ref{eq:mobility} from equilibrium MD via Eqs~\ref{eq:m11}-\ref{eq:c11}, \ref{eq:m12}-\ref{eq:c12}, \ref{eq:m21}-\ref{eq:c21} and \ref{eq:m22}-\ref{eq:c22}, which are all of the form:
\begin{align}
\mathcal{M}_{kl}^{GK}(z) &= \frac{ V  }{ k_B T } \lim_{t\to\infty} I_{kl}(t,z)
 \; ,
\label{eq:mklGK}
\end{align}
where
\begin{align}
I_{kl}(t,z)  = \int_{0}^{t} {\rm d}t'\, C_{kl}(t',z) 
 \; .
\label{eq:defIkl}
\end{align}
Computing these correlation functions remains however a great computational challenge, since it requires both (a) a fine sampling along the $z$ axis to obtain the mobility profiles and (b) a good convergence of their integral $I_{kl}(t,z)$ at long times. The former aspect renders the sampling with histograms of small bin width ($\Delta z$) difficult due to the small number of particles in each bin (with a variance diverging as $1/\Delta z$), while the latter is an even stronger requirement than a good convergence of the correlation functions themselves. In addition, the combination of both constraints result in a large memory requirements to store the correlation functions as a function of position and time with fine sampling (small bin width and time interval), and in a long simulation time to reach convergence.

Here we address both issues. Firstly, we introduce an improved estimator of the local/global correlation functions, which makes use not only of the positions and velocities of the particles, but also of the forces acting on them. Secondly, the way of computing the correlation function might be improved by using averaging algorithms such as adjustable frequency sampling\cite{frenkelUnderstandingMolecularSimulation2001}, cepstral analysis\cite{ercole_accurate_2017,baroni_heat_2018} or multiple-$\tau$ correlator \cite{ramirezEfficientFlyCalculation2010}. We have followed a different route, namely to avoid the sampling of the time-correlation functions by deriving the ``Einstein'' counterparts of the Green-Kubo expressions, \emph{i.e.} using the time-integrated currents instead of the current themselves, as proposed for the case of Poiseuille flow in Ref.~\cite{agnihotriDisplacementsMeanSquaredDisplacements2014}. The improved sampling in space and time are described in Sections~\ref{sec:forces} and~\ref{sec:einstein}, respectively.

\subsection{Sampling space: force-based estimators}
\label{sec:forces}

The sampling of correlation functions between local and global fluxes is most naturally performed by introducing bins of finite width $\Delta z$, which amounts to replacing the Dirac delta functions in the instantaneous fluxes defined by Eqs.~\ref{eq:q_z}  and~\ref{eq:j_z} by rectangular functions of width $\Delta z$ around the position $z$. As mentioned above, this strategy is plagued by a diverging variance as the bin width vanishes (fine sampling) because the estimates fluctuate between 0 (for empty bins) and occasional large values. 
Inspired by the recent development of improved estimators for the computation of local properties such as number, charge or polarization densities\cite{borgisComputationPairDistribution2013,schultzReformulationEnsembleAverages2016,delasherasBetterCountingDensity2018,colesComputingThreedimensionalDensities2019,schultzAlternativesConventionalEnsemble2019} (some of these methods based on earlier developments for Quantum Monte Carlo~\cite{assaraf_zero-variance_1999,assarafImprovedMonteCarlo2007,toulouseZerovarianceZerobiasQuantum2007}), we introduce here an alternative sampling scheme which does not involve bins (even though the computation for various positions also results in a discretization of space), but makes use of the force acting on the particles, in addition to their positions and velocities. To the best of our knowledge, such an approach has never been considered previously to sample time-correlation functions involving local fluxes (and in turn, local mobilities as discussed in the previous section).

\subsubsection{Force sampling and mixed estimators: Density}
\label{sec:forces:density}

In order to introduce the force sampling approach, we first consider a simpler
quantity, namely the local density (see Eqs.~\ref{eq:denstot}
and~\ref{eq:dens}). The canonical average of an observable $\mathcal{O}$
involves an integral over phase space with the Boltzmann weight:
\begin{align}
\left\langle \mathcal{O} \right\rangle &=
\frac{1}{\mathcal{Z}} \int \mathcal{O}({\bf r}^N, {\bf p}^N) e^{-\beta \mathcal{H}({\bf r}^N, {\bf p}^N)}
\, {\rm d}{\bf r}^N {\rm d}{\bf p}^N
\label{eq:observable}
\end{align}
where $\beta=1/k_BT$, ${\bf r}^N$ and ${\bf p}^N$ are the position and momenta of the $N$ particles,
$\mathcal{H}=\mathcal{U}({\bf r}^N)+\mathcal{K}({\bf p}^N)$ is the Hamiltonian of
the system (sum of potential energy $\mathcal{U}$ and kinetic energy
$\mathcal{K}$), and the normalization factor $\mathcal{Z}$ is the partition
function. With the specific choice $\mathcal{O}(z)=\frac{1}{S}\sum_{i=1}^N
\delta(z_i-z)$, and noting that the gradient of the Boltzmann weight with
respect to $z_i$ is $\beta f_{z,i}e^{-\beta \mathcal{H}}$ with $f_{z,i}$ the
$z$-component of the force acting on particle $i$, one obtains that the gradient
of the density with respect to $z$ is given by:
\begin{equation}
\label{eq:drhodz}
\frac{{\rm d}\rho (z)}{{\rm d} z} = \frac{\beta}{S}\left\langle \sum_{i=1}^N f_{z,i} \delta(z_i-z) \right\rangle
= \beta f_z(z)
\; .
\end{equation}

This means that, up to a factor $\beta$, the number density can be obtained as
the integral (with respect to $z$) of the force density $f_z(z)$. However, 
straightforward integration of the right-hand side, resulting in Heaviside
functions instead of Dirac deltas, may lead to a spurious non-zero density in
regions where no particles are present (inside the solid 
walls)\cite{delasherasBetterCountingDensity2018,purohit_force-sampling_2019}.

As an improvement with respect to simple integration of Eq.~\ref{eq:drhodz},
which we will later use in the following extension to correlation functions,
we propose a combination of estimators involving both the force and number 
densities (further details can be found in Appendix~\ref{app:forces}).
To that end, we introduce a weight function $w_N$,
discussed below, and define the following estimator:
\begin{align}
\tilde{\rho}(z) = &
\frac{1}{S} \left\langle \sum_{i=1}^N w_N(z_i-z) \right\rangle
-  \frac{\beta}{S} \left\langle \sum_{i=1}^N f_{z,i} w_f(z_i-z) \right\rangle 
\; ,
\label{eq:rho_fsampl}
\end{align}
where the weight function of the force density is related to that of the 
number density as:
\begin{equation}
\label{eq:weightforce}
w_f(z) = \Theta(z) - W_N(z) 
\end{equation}
with $\Theta$ the Heaviside function and $W_N$ an antiderivative of $w_N$
such that $w_f$ vanishes when $|z|$ is large. This sets some constraints
on the choice of the weight $w_N$, which can be seen as a coarse-graining
kernel for the contribution of each particle to the number density.
This function should therefore vanish beyond a coarse-graining length $\xi$.
The constraint on $w_f$ further imposes that the integral of $w_N$ is equal
to 1. While others are possible, we make here the simple choice
of a triangular kernel:
\begin{align}
w_N(z) =  \left\lbrace
\begin{array}{l}
 (\xi-|z|)/\xi^2 \; \textrm{for}\; z\in[- \xi , \xi] \\ 
 \textrm{0 otherwise}
 \end{array}  
 \right.
\label{eq:tria}
\end{align}
with dimension of a reciprocal length, from which $w_f$, which is dimensionless,
is easily determined.
We have also considered rectangular or trigonometric kernels, with similar
results. 
A more important point is the choice of the length $\xi$, on which
the final estimate depends: For $\xi\to0$ the only contribution to
$\tilde{\rho}$ is that of the number density, since in this limit
$w_N(z)\sim\delta(z)$ and $w_f(z)\sim0$, while for $\xi\to\infty$
only the integral of the force density (corresponding to Eq.~\ref{eq:drhodz}) 
contributes, since $w_N(z)\sim0$ and $w_f(z)\sim \Theta(z)$.
We have found that a value of $\xi=0.1\sigma$, with $\sigma$ the molecular
diameter, provides a good compromise between the two estimates, 
with the benefit of reduced variance of force sampling, while mitigating
the artefact of non-vanishing density in regions where no particles are present.
In practice, the estimator introduced in Eq.~\ref{eq:rho_fsampl}
can be computed efficiently by convoluting (a posteriori) the
histogram-based estimators of the number and force densities with their
corresponding weight functions (Eqs.~\ref{eq:weightforce} and~\ref{eq:tria}), as:
\begin{align}
\tilde{\rho}(z) = &\int_0^H {\rm d}z'\, 
\left[  w_N(z'-z) \frac{1}{S} \left\langle \sum_{i=1}^N \delta(z_i-z') \right\rangle
\right. 
\nonumber \\
& \left.  
\hskip 1.5cm - w_f(z'-z)  \frac{\beta}{S} \left\langle \sum_{i=1}^N f_{z,i} \delta(z_i-z') \right\rangle 
\right] 
\nonumber \\
= &\int_0^H {\rm d}z'\, \left[ w_N(z'-z) \rho (z') -  w_f(z'-z) \beta f_z(z') \right]
\nonumber \\
= & (w_N\ast \rho)(z) + \beta (w_f \ast f_z)(z)
\;, 
\label{eq:rho_fsampl2}
\end{align} 

where in the last line $\ast$ denotes the convolution product.
We finally note that, even though the main novelty of the present work is to apply this
force sampling strategy to local transport properties, as described in the next
section, the proposed combination of estimators is, to the best of our knowledge, 
also new for the density.

\subsubsection{Time correlation functions}
\label{sec:forces:tcf}

Since the time-correlation $C_{kl}(t,z)$ are also defined as canonical averages
of observables involving Dirac delta functions, we propose to apply the same
strategy to obtain estimators with reduced variance as the one above for the
number density. We illustrate this idea on the particular case of 
$C_{11}(t,z)=C_{qQ}(t,z)$, but the extension to other correlation functions 
is straightforward.
Starting from the definitions Eqs.~\ref{eq:c11}, \ref{eq:q_z} and \ref{eq:Qoft},
this correlation function is the ensemble average: 
\begin{align}
C_{qQ}(t,z) &= \left\langle 
Q(0) \frac{H}{N} \sum \limits_{i=1}^{N} v_{x,i}(t) \delta (z_i(t)-z) 
\right\rangle
\; .
\label{eq:c112}
\end{align}
The procedure leading to the mixed estimator for the density presented in the
previous section can be followed, introducing the same weight functions
$w_N$ and $w_f$, as well as a new force-weighted observable namely: 
\begin{equation}
F_{qQ}(t,z) = \left\langle 
Q(0) \frac{H}{N} \sum \limits_{i=1}^{N} v_{x,i}(t) f_{z,i}(t)\delta (z_i(t)-z) 
\right\rangle
\; ,
\label{eq:c_f}
\end{equation}
and form the mixed estimator:
\begin{align}
\tilde{C}_{qQ}(t,z) 
= &\int_0^H {\rm d}z'\, \left[ w_N(z'-z) C_{qQ}(t,z')  -  w_f(z'-z) \beta
F_{qQ}(t,z') \right]
\nonumber \\
= & (w_N*C_{qQ})(t,z)  +  \beta (w_f *F_{qQ})(t,z)
\; ,
\label{eq:c_f_sampl}
\end{align}
where the convolution products are in space only, not time.
Note that the average defining $C_{qQ}(t,z)$ is taken over the canonical distribution of 
initial conditions, \emph{i.e.} with a Boltzmann weight corresponding to the point
of phase space at time 0, while some observables are considered at the
subsequent time $t$. The derivation leading to the exact result for the density
(where all microscopic observables are considered at the same time), which
involves an integration by parts over the initial positions $z_i(0)$, introduces
an additional term, namely 
$\left\langle Q(0) \frac{H}{N} \sum_{i=1}^N \frac{\partial v_{x,i}(t) }{\partial
z_i(0)} w_f(z_i-z) \right\rangle$,
which involves the derivative of the $x$ component of the velocity at time
$t$ with respect to the initial position in the $z$ direction.
Although this term might not vanish in principle (we were not able to derive
that it does), it would be very difficult to evaluate from the trajectories.
However, we observe numerically that the estimator defined by
Eq.~\ref{eq:c_f_sampl}, which neglects it, provides the same result
as histograms based on Eq.~\ref{eq:c112}, with a lower variance.
Such a cancellation of this term (which may not be exact) probably
arises from the symmetry of the system and the considered observables,
and does not hold a priori for arbitrary time correlation functions.
To the best of our knowledge, the present work is the first to extend
the idea of using force-based estimators to time-correlation functions
(hence local mobilities). 
For the other correlation functions $C_{kl}(t,z)$, the same weights
are used in Eq.~\ref{eq:c_f_sampl}, with the analogs of Eq.~\ref{eq:c_f}
for the corresponding force-based estimators.

\subsection{Sampling time: integrated fluxes}
\label{sec:einstein}

As mentioned above, the second challenge is to converge, for each position $z$,
the integral of the time correlation functions $C_{kl}(t,z)$, which is even more
difficult than to converge the $C_{kl}(t,z)$ themselves. The direct approach
requires both a fine sampling of short times, where the function varies
significantly, possibly with cancellations between positive and negative
contributions, and accurate estimates of the decay at long times, because
even small values may lead to a non-negligible contribution to the integral.
We therefore follow another route, which is to consider the Einstein-Helfand
counterpart of the relevant Green-Kubo expressions 
Eqs~\ref{eq:m11}, \ref{eq:m12}, \ref{eq:m21} and~\ref{eq:c22},
\emph{i.e.} using the displacements instead of the velocities, as proposed for
the case of Poiseuille flow in Ref.~\cite{agnihotriDisplacementsMeanSquaredDisplacements2014}.

In the well-known case of the diffusion of a particle, the
diffusion can be expressed either as the integral of the velocity auto-correlation
function (Green-Kubo) or as the slope of the mean-square displacement at long timescales,
\emph{i.e.} the product of the time integral of the velocity with itself
(Einstein)\cite{hansenTheorySimpleLiquids2006}. Similarly, in the present case one can derive the following alternative expression for the integral of any cross-correlation function
$C_{xY}(t,z)$ of a local flux $x\in \lbrace  q, j_\nu \rbrace$ and
global flux $Y \in \lbrace  Q, J_\nu \rbrace$:
\begin{equation}
\int\limits_0^{+\infty} {\rm d}t\, C_{xY}(t,z) = \lim_{t \rightarrow +\infty} 
\frac {K_{xY} (t,z)}{2t} 
\; , 
\label{eq:MSDsampl}
\end{equation}
where
\begin{align}
K_{xY} (t,z) &= \left\langle \int_0^{t} {\rm d}t''\, x(z,t'') \int_0^{t} {\rm d}t' Y(t') \right\rangle  
\; .
\label{eq:Kxy}
\end{align}
The relevant combination $K_{kl}(t,z)$ of these terms can be used to compute the
integrals of the corresponding correlation functions $C_{kl}(t,z)$,
with $\lbrace  k, l \rbrace\in \lbrace  1, 2 \rbrace$, as previously,
as well as the mobility coefficients, computed via this Einstein-Helfand route
\begin{equation}
\mathcal{M}_{kl}^{EH}(z)=\frac{V}{k_B T} 
\lim_{t \rightarrow +\infty} \frac{K_{kl}(t,z)}{2t} 
\; . 
\label{eq:mklEH}
\end{equation}
This formulation of the mobility is more efficient computationally
than integrating the time correlation functions. Indeed, both methods require
evaluating quantities (currents or their running time integrals) at discrete 
times $t=n t_{sampl}$, but the sampling time $t_{sampl}$ can be much longer
than with the correlation functions because one only needs the linear behaviour of $K_{kl}(t,z)$ at long timescales and not to resolve the variations at all time to compute the
integrals. This in turn allows to significantly decrease the computational cost 
and the memory footprint of the corresponding arrays 
-- which becomes a limiting factor for fine sampling of space (large number of $z$ values).
In practice, the terms in Eq.~\ref{eq:Kxy}
of the form $\int \limits_0^{t} dt'' x (z,t'')$ are computed on the fly 
by integrating numerically the local currents, while the $\int  \limits_0^{t} dt' Y (t') $ 
can be computed with the atom displacements. 
This last contribution requires to properly take periodic boundary conditions into account by ``unfolding'' trajectories, as emphasized in Ref.~\cite{viscardy_transport_2007} for the computation of viscosity using the Helfand-Einstein approach.
Finally, we also use the same approach for the mixed estimator involving
the forces, introduced in the previous section (see Eq.~\ref{eq:c_f_sampl})
to obtain the corresponding expressions for the $K_{kl}(t,z)$,
noted $\tilde{K}_{kl}(t,z)$.

\section{Simulation details}
\label{sec:simulations}

\subsection{System}
\label{sec:sys}

We consider a binary fluid of solute $A$ and solvent $B$, confined between two walls described by explicit particles $W$ on a square lattice. All particles interact via the Lennard-Jones (LJ) potential
\begin{equation}
V_{ij}(r) = 4 \epsilon_{ij} \left[ \left(\frac{\sigma_{ij}}{r}\right)^{12} -  \left(\frac{\sigma_{ij}}{r}\right)^{6} \right]
\; ,
\end{equation}
where $\epsilon_{ij}$ is the interaction strength, $\sigma_{ij}$ is the particle
diameter and $r$ the distance between two particles of types $i$ and $j$. In
order to keep the system as simple as possible, while keeping the differential
affinity of the solute and solvent for the wall necessary to induce
diffusio-osmotic flows, we take all diameters to be equal ($\sigma_{ij}=\sigma$
for $\{i,j\}\in\{A,B,W\}$) and all interaction strengths except that between the
solute and the wall to be equal: $\epsilon_{ij}=\epsilon$, except
$\epsilon_{AW}=1.2\epsilon$, \emph{i.e.} a stronger affinity of the solute for
the wall compared to the solvent. In addition, we consider equal masses $m$ for
solute and solvent. In the following, we will express all quantities in L.J.
units, \emph{i.e.} taking $\epsilon=1$, $\sigma=1$ and $m=1$. 

\begin{figure}[ht!]
\centering
\includegraphics[width=0.6\columnwidth]{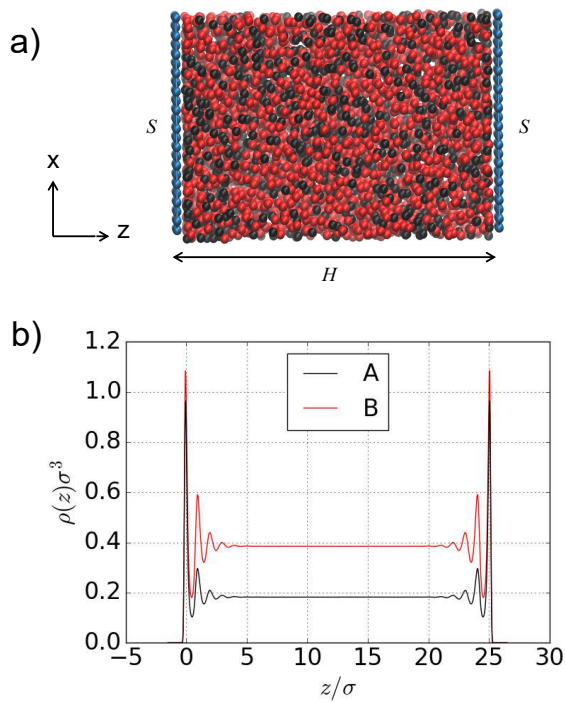}
\caption{(a) Snapshot of the simulated system, a binary fluid confined between walls separated by a distance $H$ (and a lateral surface area $S$ of the simulation box, even though periodic boundary conditions result in infinite walls).
(b) Density profiles (see Eq.~\ref{eq:dens}) for solute A and solvent B.
}
\label{fig:system}
\end{figure}

Each wall is described by a single plane of 288 fixed $W$ particles on a square
lattice, with a unit cell comprising two particles and a lattice parameter of
$\sqrt{2}\sigma$ (hence a distance $\sigma$ between particles). The box
dimensions in the $x$ and $y$ directions are
$S=12\sqrt{2}\sigma\times12\sqrt{2}\sigma$, \emph{i.e.} $12\times12$ unit cells
for each wall. The walls are separated by a distance $H+2\sigma$, where $H = 25$
(in L.J. units) is the distance between the first layers of fluid adsorbed on each
wall (see Figure~\ref{fig:system}). Periodic boundary conditions are used in the
$x$ and $y$ directions only. The fluid consists of 1440 solute and 2880 solvent
particles. Together with the dimensions of the box in the $x$ and $y$
directions, this corresponds to a total fluid density $\rho_0={N}/{V}=0.6$ (and
a reduced solute concentration ${N_A}/{V} = 0.2$). At this bulk density and the
chosen reduced temperature $T^*=k_B T/\epsilon = 1.3$, this binary mixture is
fluid\cite{trokhymchukComputerSimulationsLiquid1999}. Initial configurations are
generated by placing randomly the particles on a face cubic centered lattice and
assigning random velocities drawn from a Gaussian distribution corresponding to
the reduced temperature $T^*$. Details of the equilibration procedure are given
below.

Molecular dynamics simulations are all carried out using the LAMMPS package\cite{plimptonFastParallelAlgorithms1995}, with a time step $\Delta t=10^{-3}~t^*$ (with $t^* = \sigma \sqrt{m/\epsilon}$ the LJ time unit). Simulations are performed in the $NVT$ ensemble using a Nos\'e-Hoover thermostat (applied only along the $y$ and $z$ directions for the non-equilibrium simulation, in which a perturbation is applied along the $x$ direction), with a relaxation time of $0.1t^*$.
More details for equilibrium and non-equilibrium simulations are given in the following sections. 
In order to sample all densities and local fluxes, we consider bins along the $z$ direction of width $\Delta z = 0.014$ and sample the data every 100 steps, \emph{i.e.} $0.1t^*$. 
As illustrated in Figure~\ref{fig:system}, the considered interactions, geometry and thermodynamic conditions result in a typical structure of a fluid confined between hard walls, with a bulk-like region in which the density and composition is homogeneous and a layering of the fluid at the interface with the walls. In addition, the difference between the solute-wall and solvent-wall interactions results in a local enrichment in solute near the wall, which opens the possibility to induce a diffusio-osmotic flow.

\subsection{Non-equilibrium simulations}
\label{sec:NEMD}

For Poiseuille (pressure-driven) flows, an external force ${\bf f}^P=f^P{\bf e}_x$ in the $x$ direction is applied to each fluid particle \cite{karniadakisMicroflowsNanoflowsFundamentals2005,travisDepartureNavierStokesHydrodynamics1997,botanHydrodynamicsClayNanopores2011}. The validity of the linear response is tested by considering two forces $f^P=5.10^{-4}$ and $10^{-3}$ (in L.J. units). Since the results converge faster with the larger force, we only report the results for this case. After an equilibration for $5.10^5$ steps ($500t^*$), the properties of the steady-state system are sampled for $16000t^*$.

\begin{figure}[ht!]
\centering
\includegraphics[width=0.6\columnwidth]{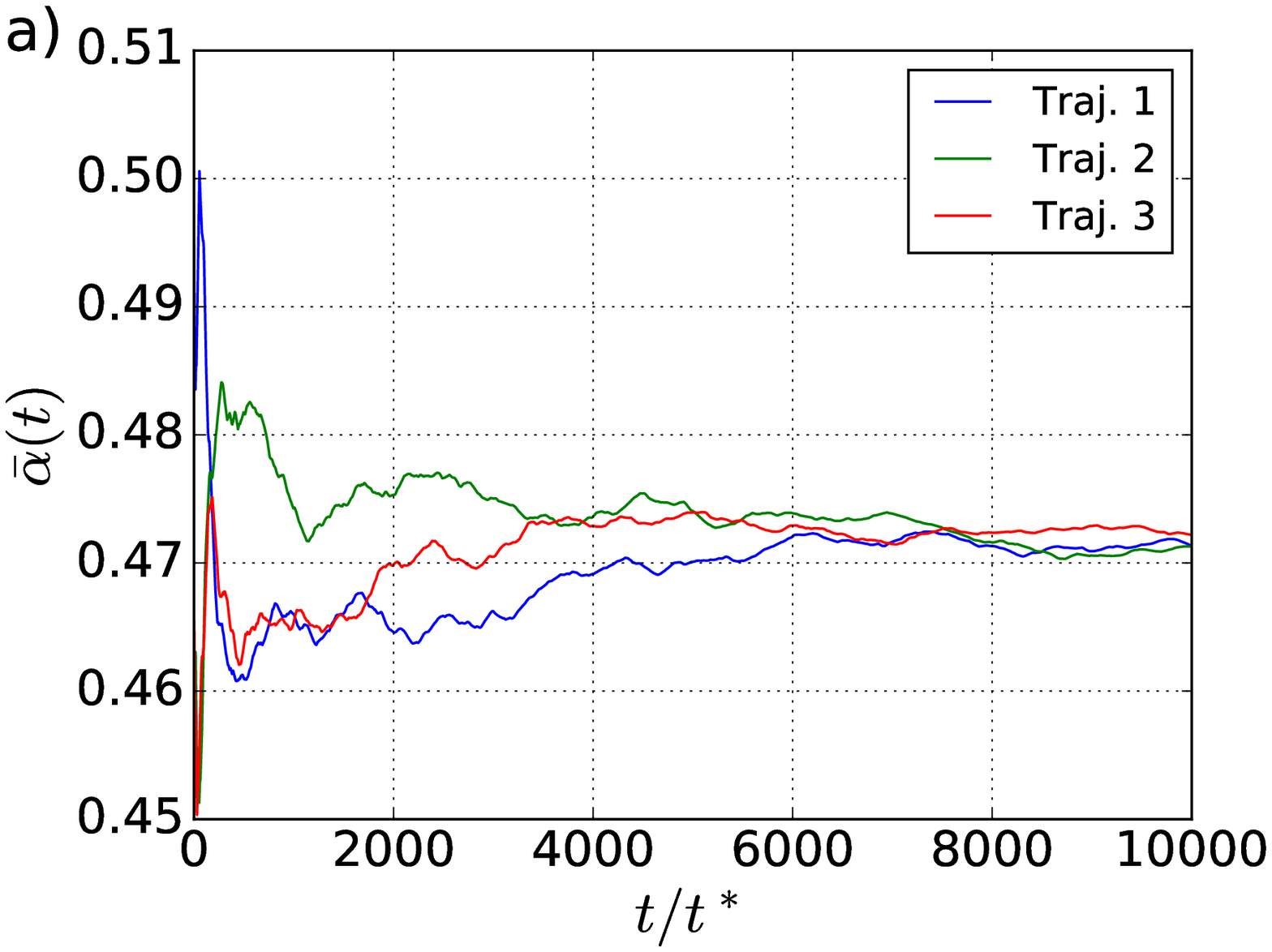}
\includegraphics[width=0.6\columnwidth]{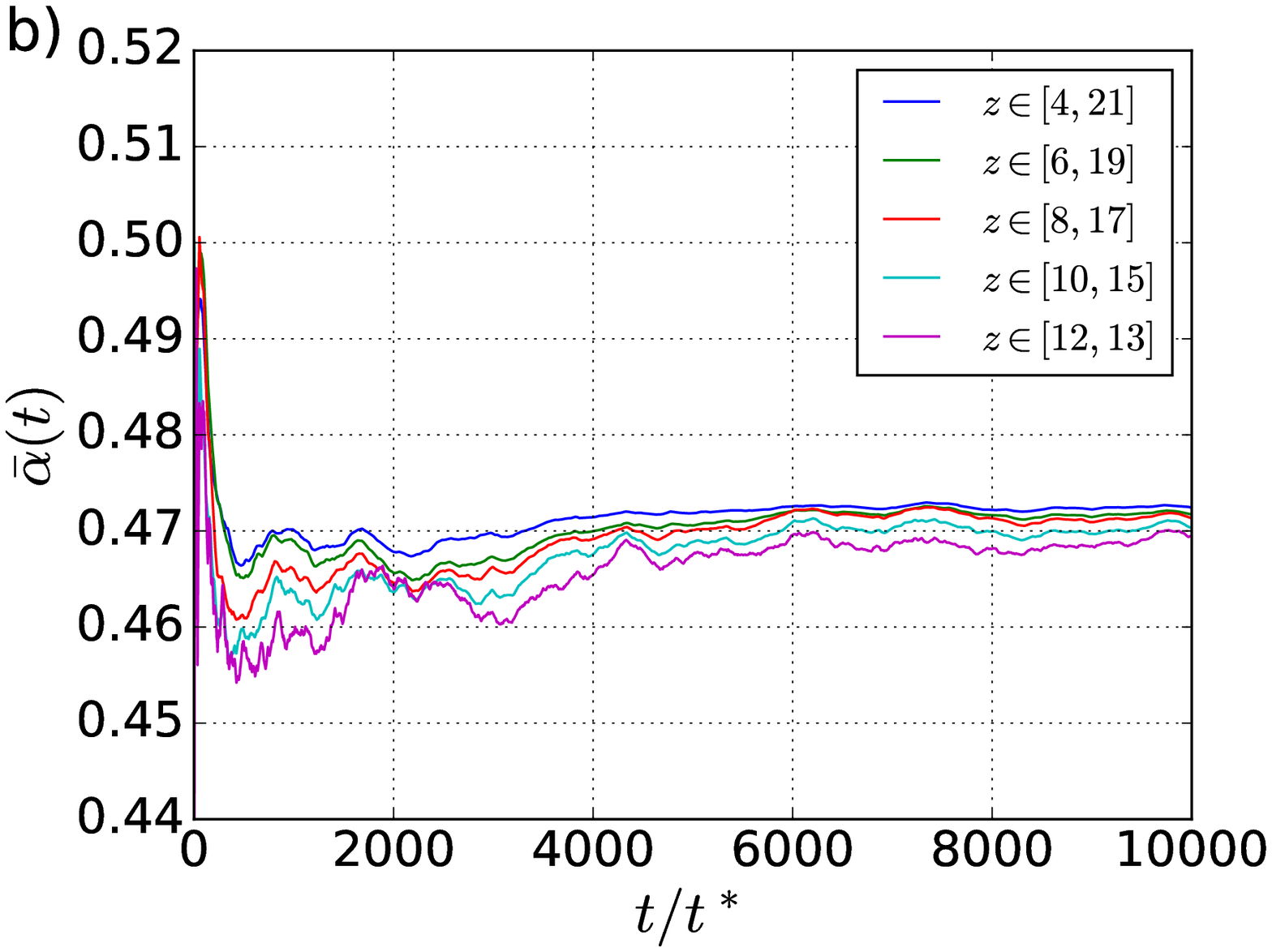}
\caption{(a) Composition of the bulk region $\overline{\alpha}$ (see Eq.~\ref{eq:alphastab}) as a
function of time for three different trajectories, with a bulk region defined
between $z=8$ and 17 L.J. units (see Fig.~\ref{fig:system} for the geometry). 
(b) $\overline{\alpha}(t)$ for a single trajectory and five definitions of the
bulk region, indicated in the panel (in L.J. units).
}
\label{fig:alpha_particles}
\end{figure}

For diffusio-osmotic flows, we use the protocol of
Ref.~\cite{yoshidaOsmoticDiffusioosmoticFlow2017}, with a force $f^\mu{\bf
e}_x$ applied on solute particles $A$ and a force $-\alpha f^\mu{\bf e}_x$
applied on solvent particles. The parameter
$\alpha$ therefore needs to be determined from prior equilibrium simulations.
Figure~\ref{fig:alpha_particles}a shows the evolution of the time dependent
estimate of the average ratio:
\begin{equation}
\overline{\alpha}(t) = \frac{1}{t} \int_0^t {\rm d}t' \, \frac{N_A^b(t')}{N_B^b(t')}
\; ,
\label{eq:alphastab}
\end{equation}
with $N_A^b$ and $N_B^b$ introduced in Section~\ref{sec:theoryfluxes},
for three different initial conditions. It illustrates the fact that an accurate
estimate of $\alpha=\lim_{t\to+\infty}\overline{\alpha}(t)$ requires a long
simulation time to converge. In addition, this quantity depends on the
definition of the ``bulk'' region in this confined system.
Figure~\ref{fig:alpha_particles}b shows the evolution of the time-dependent
estimate $\overline{\alpha}(t)$ for a single trajectory with different
boundaries ($z$ values) defining the region in which the numbers of $A$ and $B$
particles are sampled. A compromise needs to be found between too large slabs,
influenced by the inhomogeneities induced by the walls, and too small ones, in
which the smaller number of particles results in larger fluctuations. 

Based on these results, we choose in the following to define the bulk region as
comprised between $z=8$ and 17 L.J. units, and use a sampling time of $7500t^*$
(after an equilibration time of $500t^*$) without applied force to estimate
$\alpha$. This first step is carried out for every initial condition, and the corresponding value of $\alpha$ is then used for the non-equilibrium simulations on the same system. While one could in principle determine this parameter only once, and then use it for the NEMD simulations for several initial conditions, proceeding like this allows to have a single workflow. The chosen simulation times to converge the value of $\alpha$ for each trajectory result in a good consistency between them, with $\alpha=0.4733\pm0.0002$ among $N_{traj}=150$ trajectories. We then apply the perturbation described above and sample the
properties of the steady-state system during $16000t^*$ (after a further
equilibration time of $500t^*$ in the presence of the perturbation). As in the
pressure-driven case, the validity of the linear response is tested by
considering two forces $f^\mu=5.10^{-3}$ and $10^{-2}$ (in L.J. units). Since the
results converge faster with the larger force, we only report the results for
this case.

\subsection{Equilibrium simulations}

The mobility coefficients are determined from equilibrium simulations
(\emph{i.e.} in the absence of external perturbation) using Eqs.~\ref{eq:m11},
\ref{eq:m21}, \ref{eq:m12} and \ref{eq:m22}, or their extensions described in
Section~\ref{sec:improving}. The cross-correlation functions
entering in these equations are sampled from $16000t^*$-long trajectories (after
500$t^*$ of equilibration), and integrated numerically over time. The density
profile $\rho(z)$ and bulk density ratio $\alpha$ are computed simultaneously. 
In practice, the correlation functions are computed on-the-fly 
using a LAMMPS-Python interface, and a Fortran code embedded
in Python via F2Py\cite{f2py}. The results for each trajectory are then averaged
over the whole set of equilibrium trajectories. In addition, since in the
absence of external perturbation the two directions along the surface are
equivalent in the considered system, we further average the results for the
cross-correlation functions computed for the $x$ and $y$ components of the
fluxes. 

For the Green-Kubo route, the correlation functions $C_{kl}(t,z)$
and $F_{kl}(t,z)$ are computed with a sampling time 
$t_{sampl}=100 \Delta t=0.1t^*$ up to a correlation time $t^{GK}_{corr}=800 t^*$,
while for the Einstein-Helfand approach $K_{kl}(t,z)$ and corresponding 
force-weighted quantity entering in $\tilde{K}_{kl}(t,z)$ are computed
with a sampling time $t_{sampl}=1000\Delta t=t^*$ up to a correlation time 
$t^{EH}_{corr}=1000 t^*$. In all cases, the calculations are done simultaneously for 2000 $z$ values, separated by $\Delta z = 0.014$.

\section{Results}
\label{sec:results}

We now compare the various strategies detailed above to predict the mobility profiles. We first illustrate the results obtained with a fixed number of trajectories for $\mathcal{M}_{11}(z)$ and $\mathcal{M}_{12}(z)$, which correspond to Poiseuille flow and diffusio-osmosis, in Sections~\ref{sec:results:m11} and \ref{sec:results:m12}, respectively. We then analyze in more detail the efficiency by considering, for these two cases, the scaling of the standard error with the number of trajectories used to estimate the profile in Section~\ref{sec:results:efficiency}. We finally summarize all the mobility profiles corresponding to the matrix $\mathcal{M}(z)$ (see Eq.~\ref{eq:mobility}) in Section~\ref{sec:results:mall}.

For an observable $A$ (correlation function or transport coefficient as a function of time and/or position), the reported error bars correspond to the standard errors computed from independent trajectories (obtained with the same protocol but modifying the random number seed for the initial positions and initial velocities) as:
 \begin{align} 
 \sigma_{A} &= \frac{1}{\sqrt{N_{traj}(N_{traj} -1)}} \sqrt { \sum \limits_{i=1}^{N_{traj}} (A_i - \left\langle A \right\rangle)^2 }
 \label{eq:stddev}
 \end{align}
 where $N_{traj}$  is the number of trajectories, $A_i$ is the value for the $i^{\rm{th}}$ trajectory, $\left\langle A \right\rangle = \frac{1}{N_{traj}} \sum \limits_{i=1}^{N_{traj}} A_i$ is the average of $A$ over trajectories. When necessary, propagation of uncertainties is carried out with the usual formulas between independent variables (for instance when averaging in the $x$ and $y$ direction).

\subsection{Mobility coefficients from equilibrium MD: Poiseuille flow}
\label{sec:results:m11}

We first consider the mobility profile $\mathcal{M}_{11}(z)$, which corresponds to the response of the whole fluid to a pressure gradient, \textit{i.e} a Poiseuille flow, as obtained by equilibrium and non-equilibrium simulations. In all cases, we use here the same number of independent trajectories for all methods, namely $N_{traj}=150$.

\begin{figure}[ht!]
     \includegraphics[width=0.6\columnwidth]{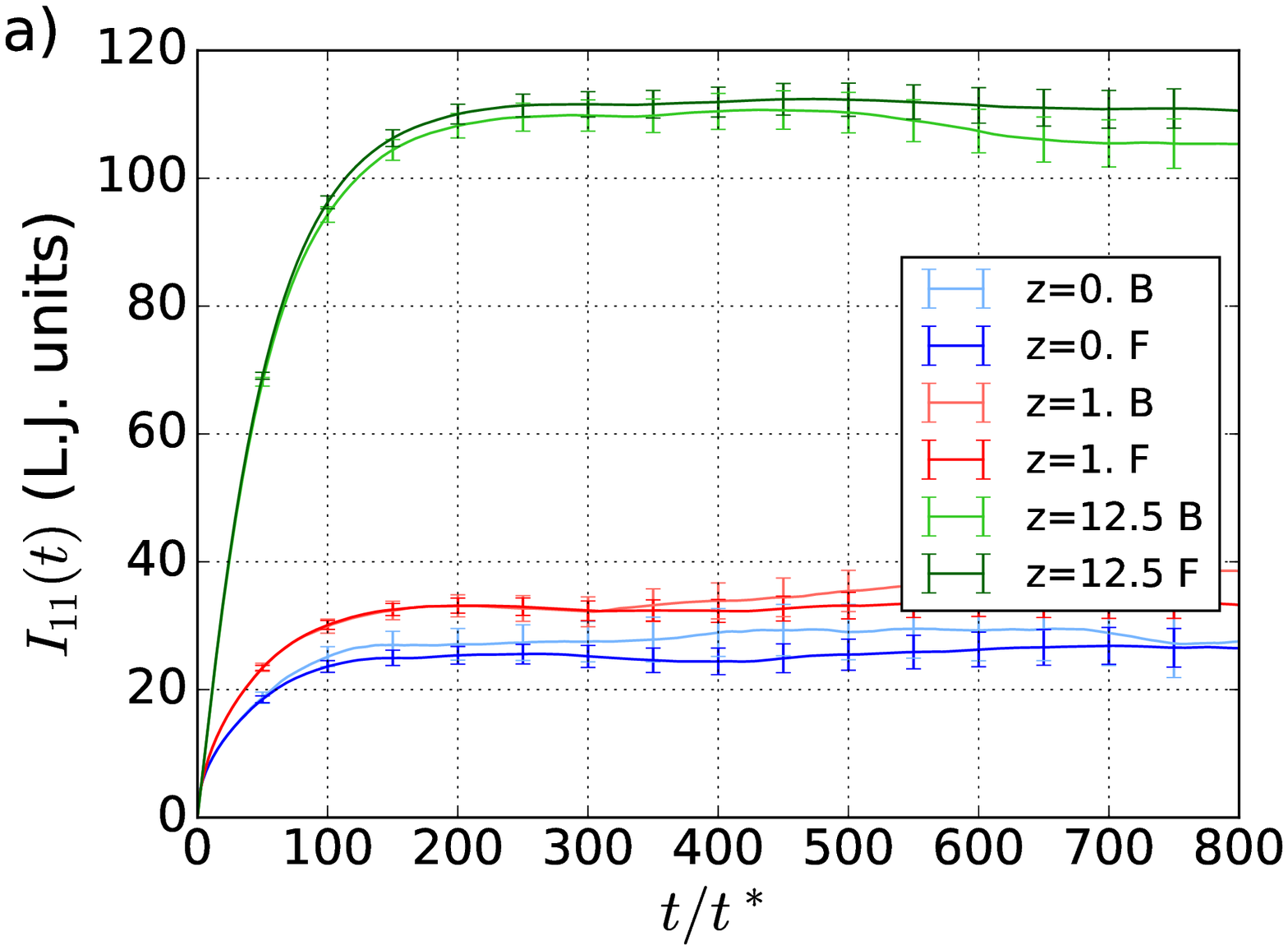}
     \includegraphics[width=0.6\columnwidth]{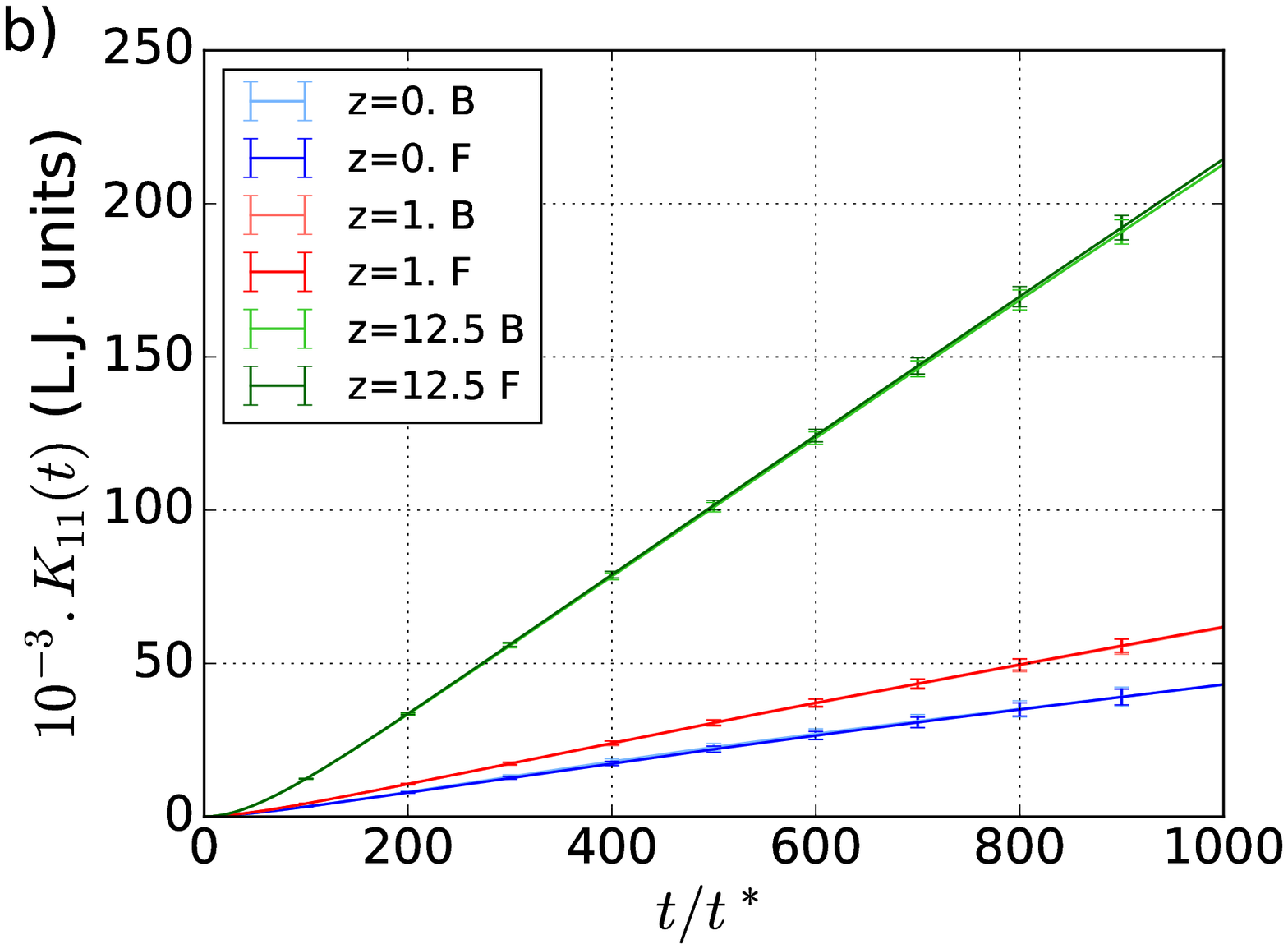}
     \includegraphics[width=0.6\columnwidth]{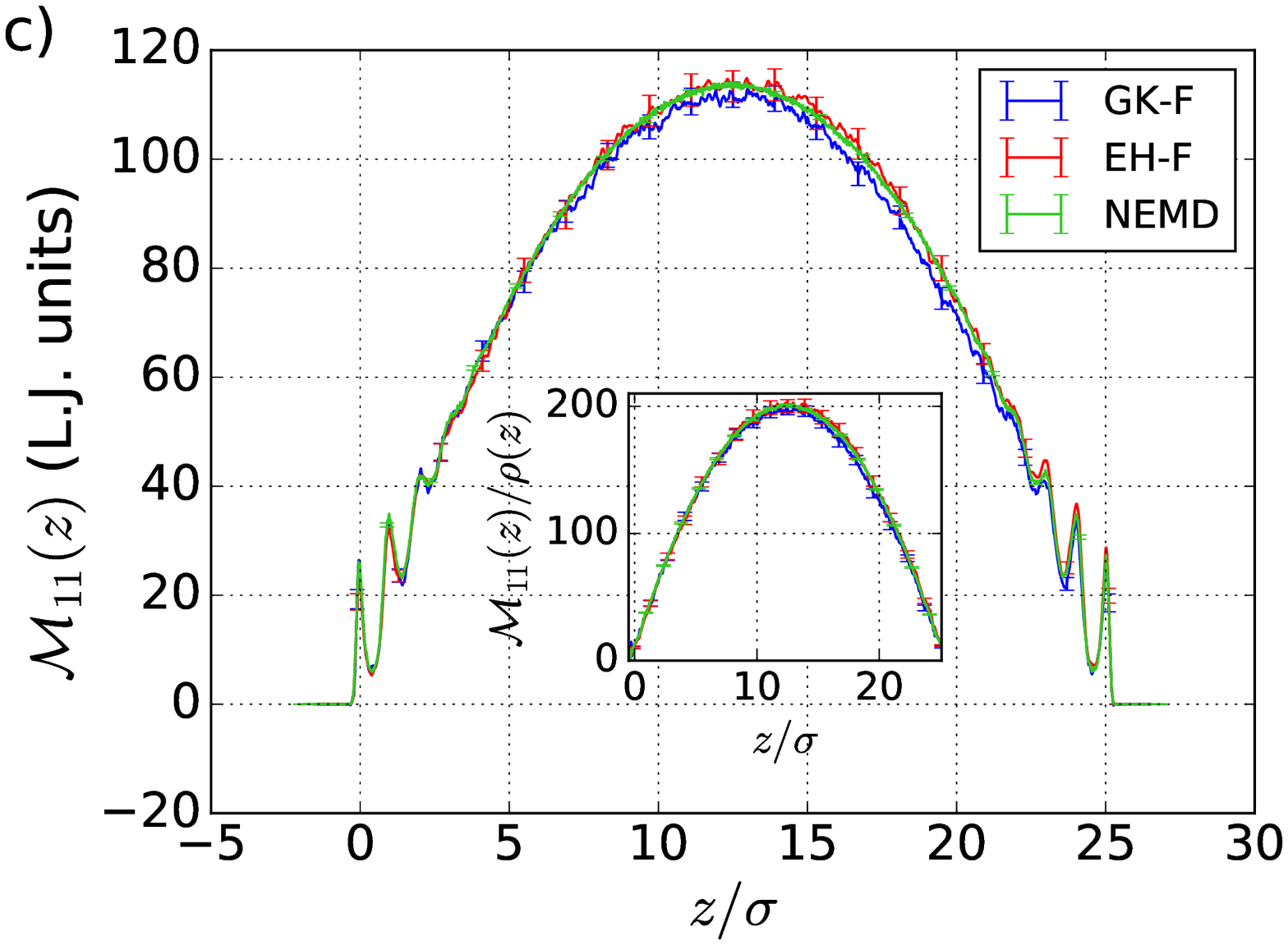}
\caption{
\textbf{Total flux induced by a pressure gradient (Poiseuille flow) from equilibrium MD}
(a) Running integral $I_{11}$ of the correlation function $C_{11}$ (see Eqs.~\ref{eq:defIkl} and~\ref{eq:c11}), which leads to the Green-Kubo (GK) estimate of the corresponding mobility profile (see Eq.~\ref{eq:mklGK}), for three positions $z=0$, 1 and 12.5 (in L.J. units). Results are shown both for the standard binning approach (B label), or using the mixed estimator involving the forces described in section~\ref{sec:forces:tcf} (F label).
(b) Correlation function $K_{11}$ of the integrated currents (see Eqs.~\ref{eq:Kxy}, with $x=q$ and $Y=Q$), which leads to the Einstein-Helfand (EH) estimate of the corresponding mobility profile (see Eq.~\ref{eq:mklEH}), for the same positions.
(c) Mobility profile $\mathcal{M}_{11}(z)$ obtained with the GK and EH routes, both with the mixed estimator involving the forces; results of non-equilibrium MD (NEMD) for an external force $f^p=10^{-3}$ (in L.J. units) is also indicated as a reference. The inset shows that $\mathcal{M}_{11}(z)/\rho(z)$ exhibits the parabolic shape expected from continuum hydrodynamics for the velocity profile (see Eq.~\ref{eq:poiseuille}).
All results are shown for the same number of independent trajectories ($N_{traj}=150$).
}
\label{fig:results:M11}
\end{figure}

The Green-Kubo approach is illustrated on Figure~\ref{fig:results:M11}a, which shows the running integral $I_{11}$ of the correlation function $C_{11}$ (see Eqs.~\ref{eq:defIkl} and~\ref{eq:c11}) for three positions $z$ across the pore, corresponding approximately to the density maxima corresponding to the first and second adsorbed layers on one wall, and to the center of the pore. The mobility is obtained from the plateau using Eq.~\ref{eq:mklGK}, which is relatively well converged despite the increase in variance at longer times (expected for an integral of a correlation function). This panel also reports the results obtained using the mixed estimator involving the force in addition to the velocities of the particles, given by Eq.~\ref{eq:c_f_sampl}. The latter is seen to result in a flatter plateau and a (slightly) lower variance on $I_{11}(t)$, which indicates that it provides the mobility with a better accuracy for the same number of trajectories (this will be discussed in more detailed in Section~\ref{sec:results:efficiency}).
Based on the results of Figure~\ref{fig:results:M11}a and the corresponding results for the other transport coefficients (in particular Figure~\ref{fig:results:M12}a below), in order to compute the mobility profile $\mathcal{M}^{GK}_{11}(z)$ we evaluate the plateau value of $I_{11}(t)$ at $t=400t^*$, which provides a good compromise between the limit converged to at larger timescales and the increasing uncertainty on the estimate as $t$ increases.

Figure~\ref{fig:results:M11}b then illustrates the Einstein-Helfand, for the same positions, by showing the correlation function $K_{11}$ of the integrated currents (see Eqs.~\ref{eq:Kxy}, with $x=q$ and $Y=Q$). In all cases one observes a linear regime after a few 100$t^*$ and the slope allows to compute the mobility via Eq.~\ref{eq:mklEH}. As for the Green-Kubo case, the mixed estimator involving the force, also shown on this panel, provides a slightly smaller variance on $K_{11}(t)$ for the same number of trajectories than the straightforward binning, but the benefit seems smaller than in the Green-Kubo case.
Based on the results of Figure~\ref{fig:results:M11}b and the corresponding results for the other transport coefficients (in particular Figure~\ref{fig:results:M12}b below), in order to compute the mobility profile $\mathcal{M}^{EH}_{11}(z)$ we evaluate the slope of $K_{11}(t)$ between $t=250t^*$ and $600t^*$.

Finally, Figure~\ref{fig:results:M11}c shows the whole mobility profile across the slit pore, obtained by the non-equilibrium, Green-Kubo and Einstein-Helfand routes. For the two equilibrium approaches, only the results with the mixed estimators using the forces are indicated. Note that for EH the reported standard errors correspond to that on the slope, computed for each trajectory from $K_{11}(t)$, and not by estimating the slope on the average $K_{11}(t)$ reported in panel b. 
It is clear from this figure that both equilibrium approaches (GK and EH) are able to reproduce the NEMD results, including the parabolic flow profile expected far from the walls and the deviations from this profile near the latter, due to the layering of the fluid (see the inset of panel \ref{fig:results:M11}c, which shows that $\mathcal{M}_{11}(z)/\rho(z)$ exhibits a parabolic shape, and Eq.~\ref{eq:poiseuille} below). One can further note that the EH approach is more accurate than the GK one: Not only are the standard errors smaller with the former than the latter, but the profile also coincides better with the reference NEMD results. This could be due to the fact that we have neglected a term in the derivation of Eq.~\ref{eq:c_f_sampl}, as explained in Section~\ref{sec:forces:tcf}, but since it is very difficult to estimate this (small) term such a hypothesis is difficult to test directly.

While it is not the purpose of the present work to analyse this profile in detail, we can note that the profile is roughly consistent with no-slip boundary conditions (vanishing velocity corresponding to the parabolic profile near the wall). This is consistent with the high density of solute particles strongly interacting with the walls. In addition, the curvature of the parabola then provides a measure of the viscosity $\eta$ of the fluid in the central region. Indeed, the steady-state solution of the Stokes equation in the case of a Poiseuille flow for a homogenous fluid with density $\rho$ and no-slip boundary conditions at the walls placed at $z=0$ and $z=H$, corresponds to the following mobility:
\begin{align} 
\mathcal{M}_{11}^{P,ns}(z) = \frac{\rho}{2\eta} \left[ \left(\frac{H}{2}\right)^2 - \left(z-\frac{H}{2}\right)^2 \right]
\; .
\label{eq:poiseuille}
\end{align}
The viscosities estimated from the curvature of the parabolic fits (in a central fluid slab of width 9$\sigma$, corresponding to the bulk region where density is homogeneous, see Section~\ref{sec:simulations}) of the mobility profiles obtained by the various MD approaches are summarized in Table~\ref{tab:poiseuille}. All estimates are in excellent agreement with the bulk viscosity of the same fluid (which is in fact a one-component fluid since the interactions between solvent and solute particles are identical) at the same density (measured in the central region of the pore) and temperature, reported in Ref.~\cite{meierTransportCoefficientsLennardJones2004}.

\begin{table}[ht!]
\begin{center}
\begin{tabular}{|c|c|}
\cline{2-2}
\multicolumn{1}{c|}{}& Viscosity $\eta$ \\
\multicolumn{1}{c|}{}& (L.J. units) \\
\hline 
Non-equilibrium        & $0.69\pm0.01$ \\
\hline 
Green-Kubo bin      & $0.70\pm0.01$ \\
Green-Kubo force    & $0.70\pm0.01$ \\ 
\hline 
Einstein-Helfand bin      & $0.69\pm0.01$ \\
Einstein-Helfand force    & $0.69\pm0.01$ \\ 
\hline 
Bulk~\cite{meierTransportCoefficientsLennardJones2004} & $0.69\pm0.02$    \\ 
\hline 
\end{tabular}
\end{center}
\caption{Dynamic viscosity $\eta$ obtained from the curvature of the MD mobility profiles $\mathcal{M}_{11}(z)$ in the central region of the pore (see Eq.~\ref{eq:poiseuille}), for the various methods. Results for the bulk viscosity of the same fluid at the same reduced temperature ($T^*=1.3$) and density (in the central region of the pore, $\rho^*\approx0.566$) are taken from Ref.~\cite{meierTransportCoefficientsLennardJones2004}.}
\label{tab:poiseuille} 
\end{table}

\subsection{Mobility coefficients from equilibrium MD: diffusio-osmosis}
\label{sec:results:m12}

\begin{figure}[ht!]
     \includegraphics[width=0.6\columnwidth]{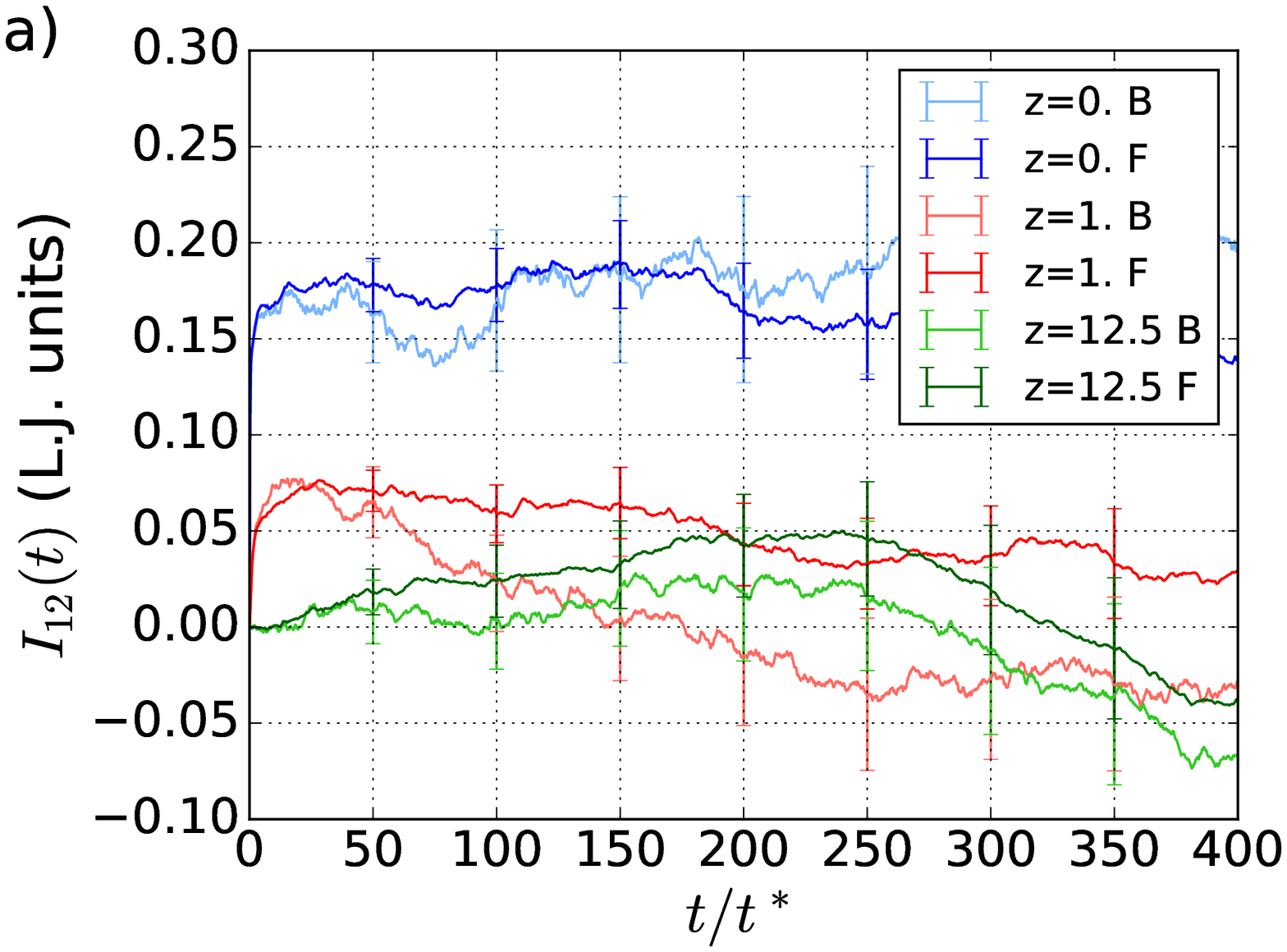}
     \includegraphics[width=0.6\columnwidth]{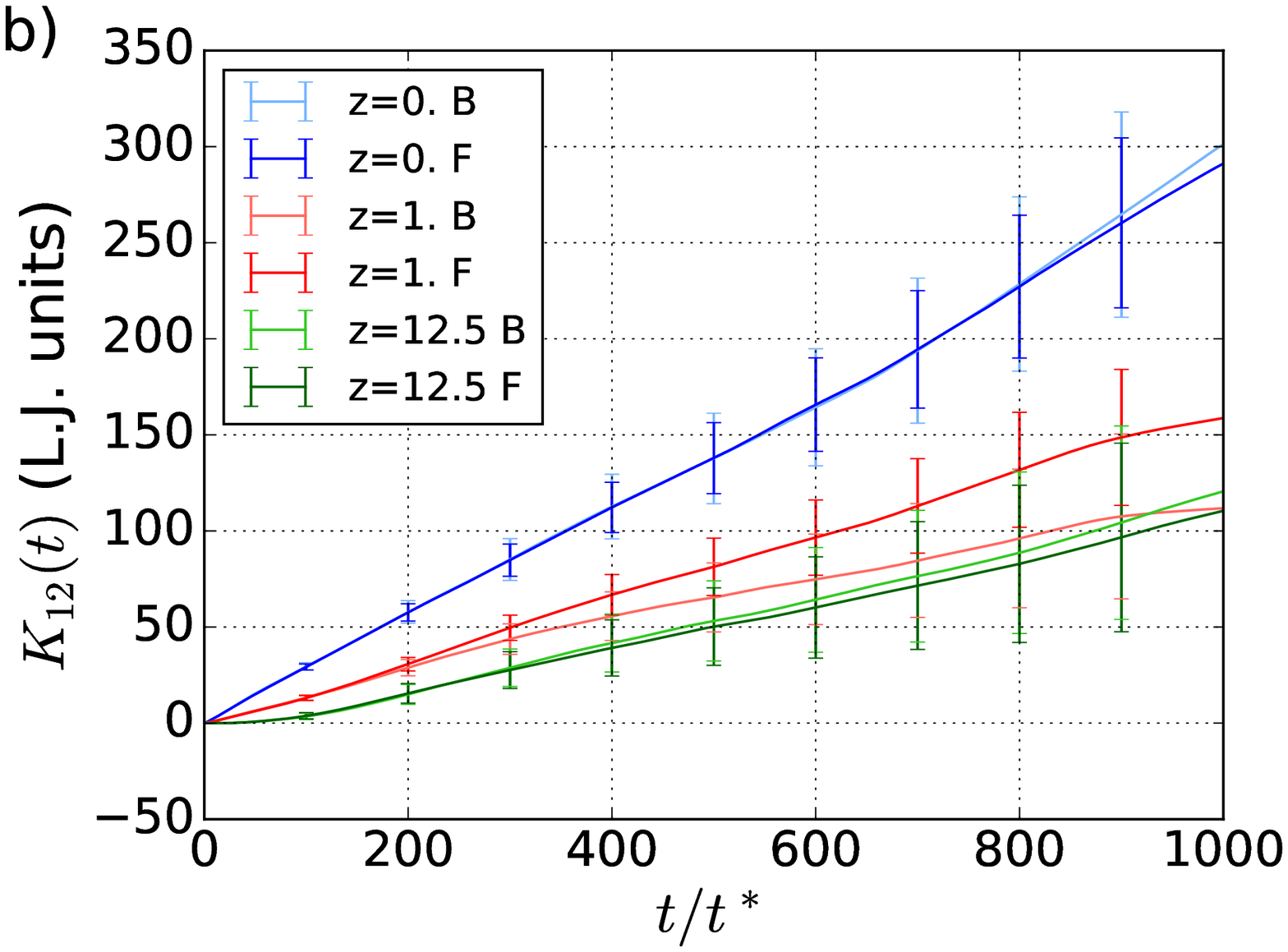}
     \includegraphics[width=0.6\columnwidth]{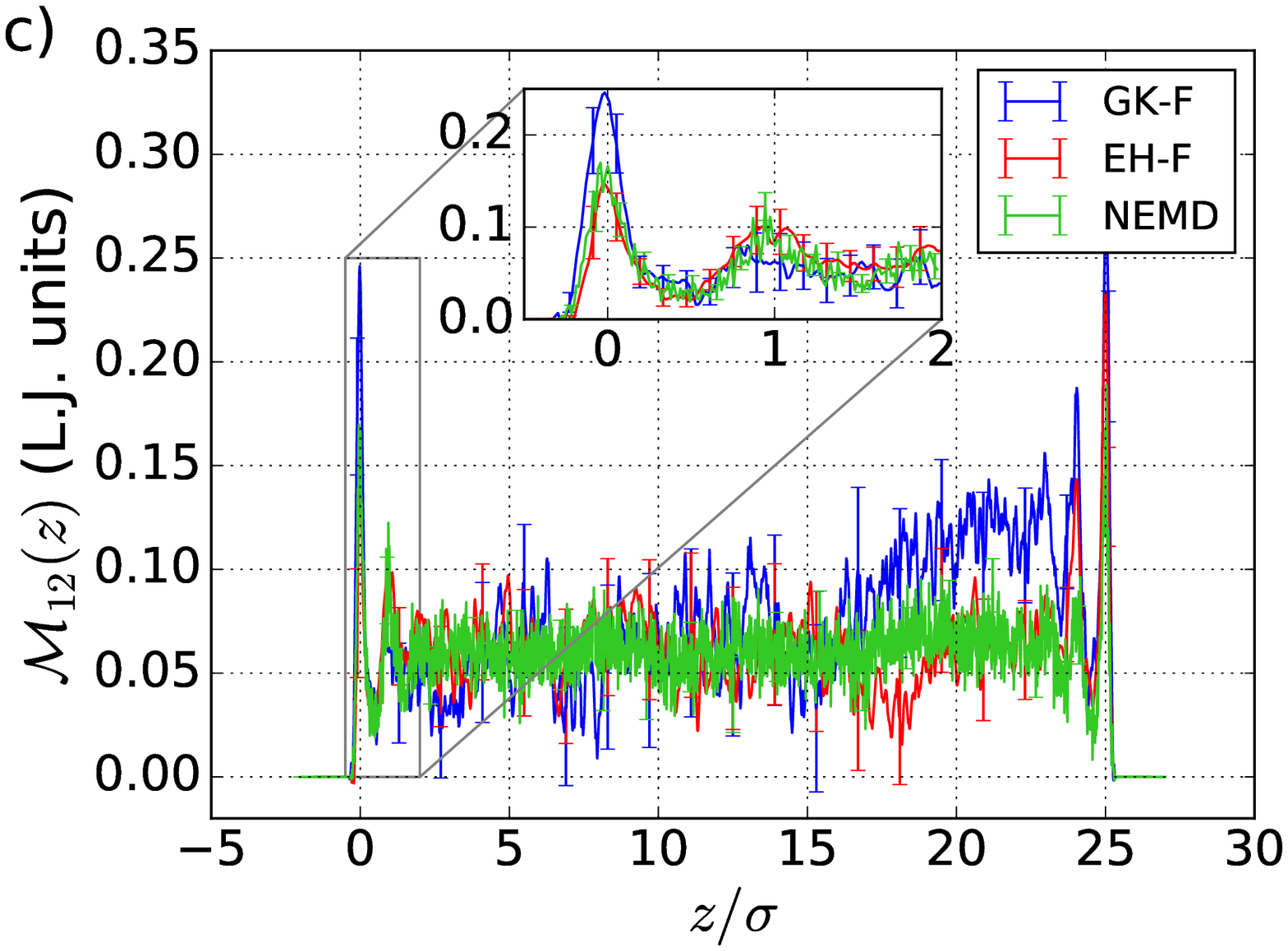}
\caption{
\textbf{Total flux induced by a chemical potential gradient (diffusio-osmosis) from equilibrium MD}
(a) Running integral $I_{12}$ of the correlation function $C_{12}$ (see Eqs.~\ref{eq:defIkl} and~\ref{eq:c12}), which leads to the Green-Kubo (GK) estimate of the corresponding mobility profile (see Eq.~\ref{eq:mklGK}), for three positions $z=0$, 1 and 12.5 (in L.J. units). Results are shown both for the standard binning approach (B label), or using the mixed estimator involving the forces described in section~\ref{sec:forces:tcf} (F label).
(b) Correlation function $K_{12}$ of the integrated currents (see Eqs.~\ref{eq:Kxy}, with $x=j_A$ or $q$ and $Y=Q$), which leads to the Einstein-Helfand (EH) estimate of the corresponding mobility profile (see Eq.~\ref{eq:mklEH}), for the same positions.
(c) Mobility profile $\mathcal{M}_{12}(z)$ obtained with the GK and EH routes, both with the mixed estimator involving the forces; results of non-equilibrium MD (NEMD) for an external force $f^\mu=10^{-2}$ (in L.J. units)  is also indicated as a reference. The inset shows the same data in the range $[-0.5,2]$.
All results are shown for the same number of independent trajectories ($N_{traj}=150$).
}
\label{fig:results:M12}
\end{figure}

We now turn to $\mathcal{M}_{12}(z)$, which corresponds to the response of the whole fluid to a chemical potential gradient, \textit{i.e} diffusio-osmosis, as obtained by equilibrium and non-equilibrium simulations. As in the previous case, we use the same number of independent trajectories ($N_{traj}=150$) for all methods. 
Figure~\ref{fig:results:M12} provides the same analysis as Figure~\ref{fig:results:M11} but for $I_{12}$ (panel a), $K_{12}$ (panel b) and the diffusio-osmotic mobility coefficient $\mathcal{M}_{12}(z)$ (panel c). As in the previous case, we compute the mobility profile $\mathcal{M}^{GK}_{12}(z)$ by evaluating the plateau value of $I_{12}(t)$ at $t=200t^*$, and $\mathcal{M}^{EH}_{12}(z)$ by evaluating the slope of $K_{12}(t)$ between $t=250t^*$ and $600t^*$.

The conclusions drawn from Figure~\ref{fig:results:M12} for diffusio-osmosis are similar to that for Poiseuille flow: The Einstein-Helfand approach is more accurate than Green-Kubo, and using the mixed estimator involving the forces in addition to the velocities reduces (slightly) the standard error on the results. An important difference with the previous case, however, is that the absolute value of the diffusio-osmotic mobility is much smaller and the relative error is much larger. This renders the convergence of the results much more difficult, \textit{i.e.} requiring a larger number of trajectories, regardless of the method used. 

Another striking difference with the response to a pressure gradient, is the fact that the diffusio-osmotic mobility profile is flat beyond a few molecular layers from the wall. This constant fluid velocity is due to the fact that there is no net force applied in the central region for this thermodynamic force. This is obvious from its microscopic mechanical analogue, which consists in applying separate forces on solvent and solutes with opposite directions and magnitudes such that they cancel in the bulk (see Section~\ref{sec:dof}).

Such a plateau of the velocity usually leads to define a diffusio-osmotic mobility $K_{DO}$ from the fluid velocity ``far'' from the walls as ${\bf v}_\infty = K_{DO} c_\infty \nabla \mu$, with $c_\infty$ the solute concentration in the bulk region\cite{marbachOsmosisMolecularInsights2019}. From the Stokes equation for a homogeneous fluid with viscosity $\eta$, and assuming no-slip boundary conditions, one can derive the following analytical expression, using the known solute concentration profile $c(z)$ from a wall located at $z=0$:\cite{yoshidaOsmoticDiffusioosmoticFlow2017}
\begin{align} 
K_{DO} &= -\frac{1}{\eta} \int_0^\infty {\rm d}z\; z \left( \frac{c(z)}{c_\infty}-1 \right) 
\; ,
\label{eq:kdo}
\end{align}
where the integral can be cut in practice at a finite distance where the concentration reaches its bulk value. This quantity is also intimately linked to the adsorption\cite{marbachOsmosisMolecularInsights2019} 
\begin{align} 
\Gamma &= \int_0^\infty {\rm d}z\; \left( \frac{c(z)}{c_\infty}-1 \right) 
\; ,
\label{eq:adsorption}
\end{align}
which characterizes the excess concentration of the solute near the wall compared to the bulk. In the case of a surface excess of solutes ($\Gamma>0$), one should observe a diffusio-osmotic velocity ${\bf v}_\infty$ in the direction opposite to $\nabla \mu$ (\textit{i.e.} $K_{DO}<0$).

\begin{table}[ht!]
\begin{center}
\begin{tabular}{|c|c|}
\cline{2-2}
\multicolumn{1}{c|}{}& Diffusio-osmotic mobility $K_{DO}$ \\
\multicolumn{1}{c|}{}& (L.J. units) \\
\hline 
Non-equilibrium        & $-0.21 \pm 0.01$ \\
\hline 
Green-Kubo bin      & $-0.21 \pm 0.01$\\
Green-Kubo force    & $-0.21 \pm 0.01$ \\ 
\hline 
Einstein-Helfand bin      & $-0.19 \pm 0.01$  \\
Einstein-Helfand force    & $-0.19 \pm 0.01$ \\ 
\hline 
Eq.~\ref{eq:kdo} & $-0.19 \pm 0.02$   \\ 
\hline 
\end{tabular}
\end{center}
\caption{
Diffusio-osmotic mobility $K_{DO}$, ratio between the plateau velocity
and the driving chemical potential gradient, or equivalently the
plateau of the mobility $\mathcal{M}_{12}(z)$ and the density $\rho(z)$
in the center of the pore. Results of the various methods are compared to the analytical prediction Eq.~\ref{eq:kdo} for fluid with uniform density and viscosity\cite{yoshidaOsmoticDiffusioosmoticFlow2017}, using the solute density profile shown in Figure~\ref{fig:system}.}
\label{tab:kdo} 
\end{table}

Table~\ref{tab:kdo} summarizes the values of $K_{DO}$ obtained by the various simulation methods, where the value of the plateau was determined as the average in the bulk region, defined as previously between $z=8$ and 17, and with Eq.~\ref{eq:kdo}. One can observe an excellent agreement between the various estimates from MD, which are also consistent with the analytical prediction Eq.~\ref{eq:kdo} (which neglects the layering of the fluid at the interface and assumes a uniform viscosity). In particular, the sign of $K_{DO}$ is negative, as expected for the present case of positive adsorption ($\Gamma=+0.36$ L.J. units), due to the stronger attraction of the wall with the solute compared to the solvent. While the discussion of this particular case is not the objective of the present work, it illustrates the ability of the equilibrium methods to compute the mobility profile corresponding to this subtle transport phenomenon.

\subsection{Efficiency of the various strategies}
\label{sec:results:efficiency}

In order to compare the various methods, we need to consider separately how their accuracy scales with the number of independent trajectories and the computer time needed to estimate the relevant properties for each trajectory. The latter are summarized in Table~\ref{tab:cputime}. Each method corresponds to a specific workflow, described in Section~\ref{sec:simulations}. 
The NEMD approaches require two separate sets of simulations: (1) applying $-\nabla_x P$ to obtain $\mathcal{M}_{11}(z)$ and $\mathcal{M}_{12}(z)$ and (2) applying $-\nabla_x \mu$ to obtain $\mathcal{M}_{21}(z)$ and $\mathcal{M}_{22}(z)$. 
In contrast, the equilibrium routes (Green-Kubo and Einstein-Helfand) provide the full mobility matrix $\mathcal{M}(z)$ simultaneously. In addition, there is virtually no additional cost to use the improved estimators using the forces. Furthermore, in the case of the applied chemical potential gradient, the NEMD approach requires prior knowledge of the bulk composition $\bar{\alpha}$ from separate equilibrium simulations (see Section~\ref{sec:NEMD}), while such additional simulations are not necessary in the equilibrium routes. This entails a non-negligible computational cost (44 of the 142 hours in Table~\ref{tab:cputime}). As a result, one should in principle compare the total cost $98+142=240$~h of NEMD per trajectory to compute the 4 profiles with the cost of the two equilibrium approaches.

\begin{table}[ht!]
\begin{center}
\begin{tabular}{|c|c|c|}
\hline 
Method & CPU time / trajectory (h) \\
\hline 
Non-equilibrium  $-\nabla_x P$  & 98 \\ 
\hline 
Non-equilibrium $-\nabla_x \mu$ & 142 \\ 
\hline 
Green-Kubo & 184 \\ 
\hline 
Einstein-Helfand  & 124 \\ 
\hline 
\end{tabular}
\end{center}
\caption{Computational time per trajectory (obtained on the same computer using 4 cores), for the various approaches considered in the present work. Each method corresponds to a specific workflow, described in Section~\ref{sec:simulations}, for each trajectory. NEMD results are given for a single magnitude of the perturbation, even though several need to be considered to check the validity of the linear response regime.  The additional cost related to the improved estimators using the force with the Green-Kubo and Einstein-Helfand approaches is negligible so that a single value is reported. 
}
\label{tab:cputime} 
\end{table}

\begin{figure}[ht!]
     \includegraphics[width=0.6\columnwidth]{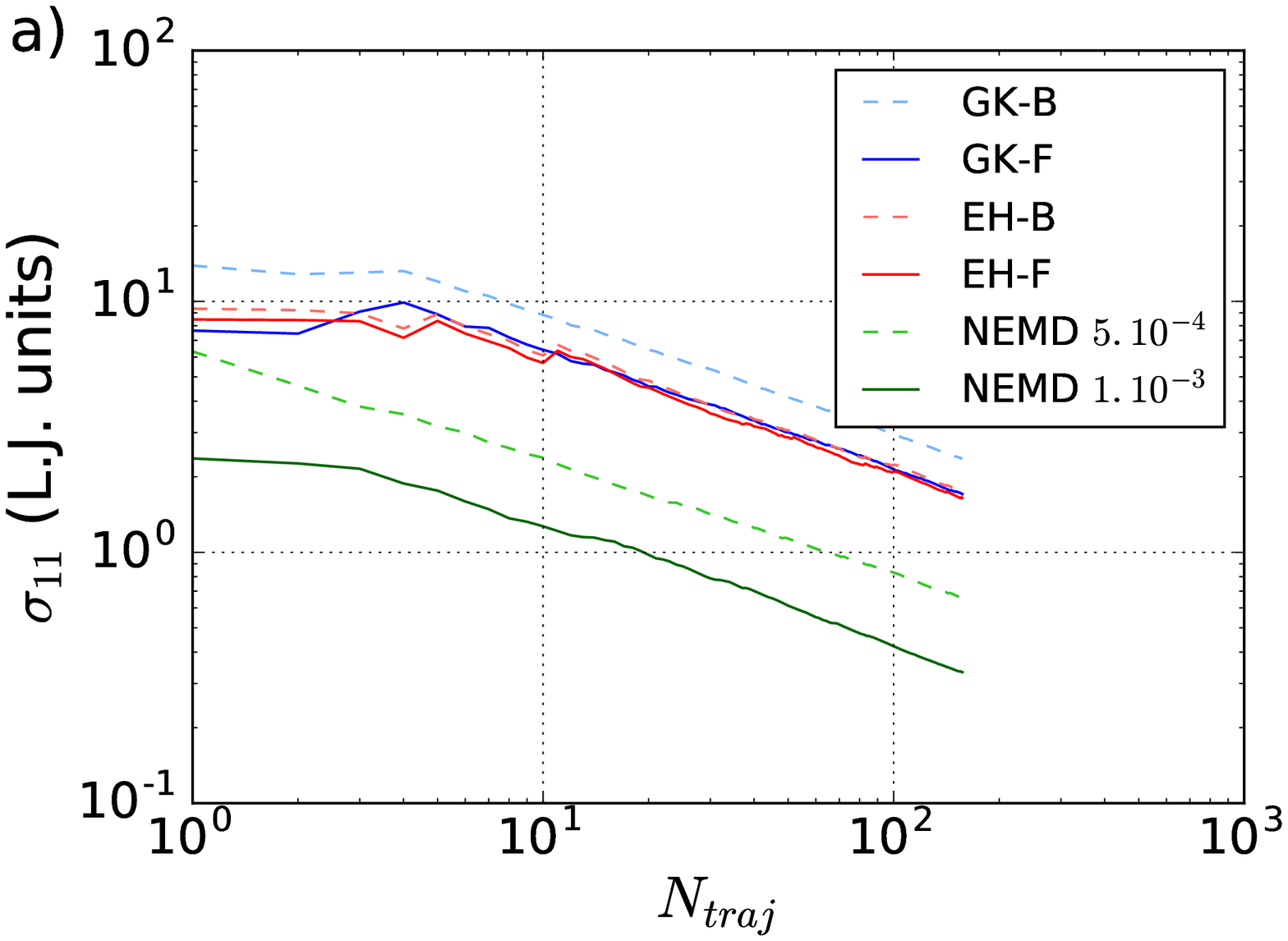}
     \includegraphics[width=0.6\columnwidth]{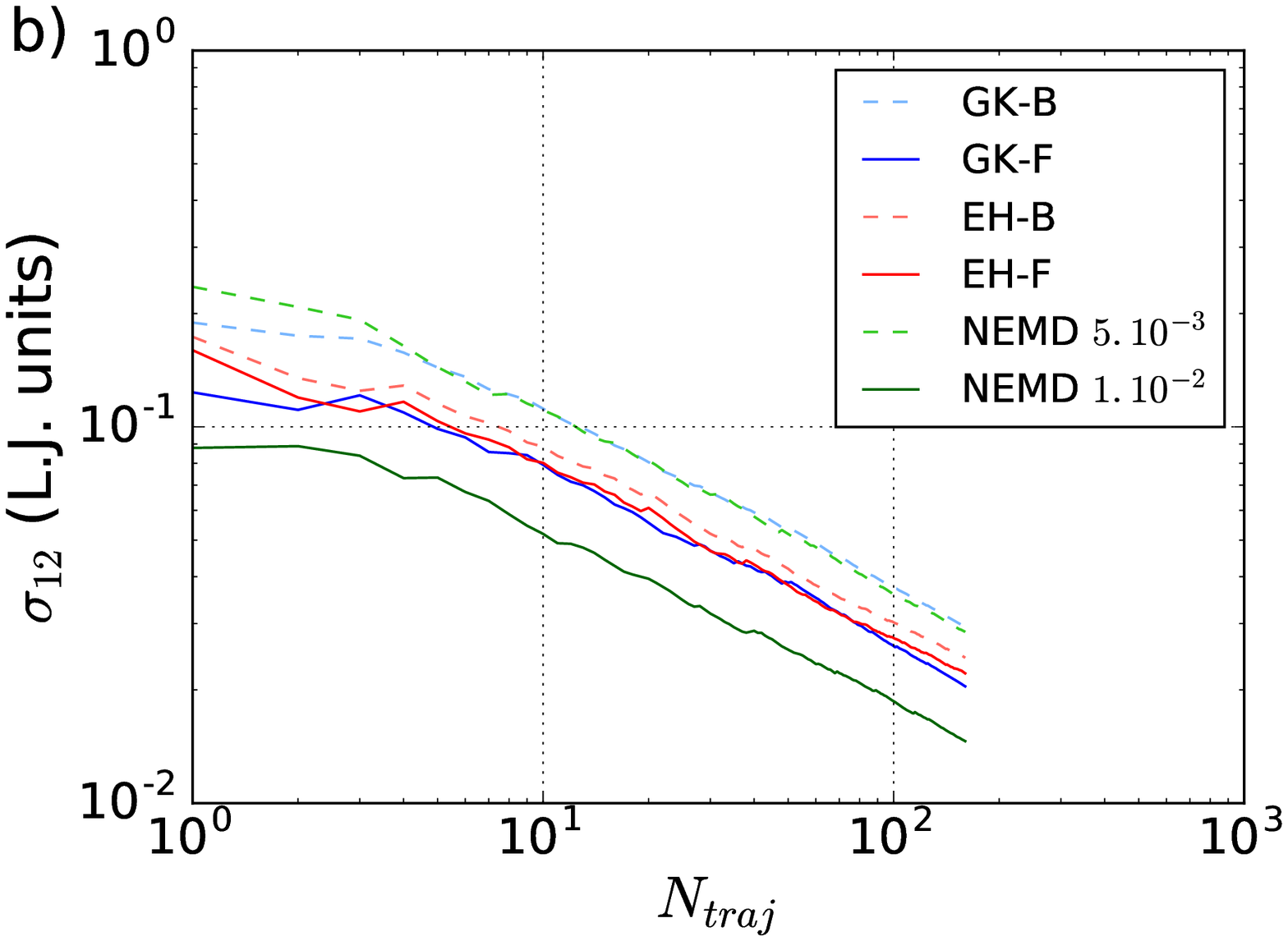}
\caption{
Standard error on the mobility averaged over all positions $z$ (see Eq.~\ref{eq:globalstddev},
as a function of the number of trajectories, with the various approaches (Green-Kubo, Einstein-Helfand, non-equilibrium) in the case of Poiseuille flow (a) and diffusio-osmosis (b).
Results for the equilibrium routes are shown both for the standard binning approach (B label), or using the mixed estimator involving the forces described in section~\ref{sec:forces:tcf} (F label). The NEMD results are shown for two sets of forces, $f^p=5.10^{-4}$ and $10^{-3}$
and $f^\mu=5.10^{-3}$ and $10^{-3}$ (L.J. units), respectively.
}
\label{fig:results:scaling}
\end{figure}

We now turn to the number of trajectory $N_{traj}$ required to achieve a given accuracy for one of the mobility profiles $\mathcal{M}_{kl}$, which also depends on the method. As a measure of this accuracy, we consider the average standard error over the profiles:
\begin{align} 
 \sigma_{kl} &= \frac{1}{N_{z}} \sum \limits_{j=1}^{N_{z}} \sigma_{\mathcal{M}_{kl}(z_j)}
 \label{eq:globalstddev}
 \end{align}
with the standard error on the mobility $\mathcal{M}_{kl}$ computed for each position $z_j$ from the ensemble of trajectories via Eq.~\ref{eq:stddev}. 
Its scaling with the number of trajectories is illustrated in Figures~\ref{fig:results:scaling}a for $\mathcal{M}_{11}$ and \ref{fig:results:scaling}b for $\mathcal{M}_{12}$. 

Despite quantitative differences for these two transport properties, due in particular to the different order of magnitude of these quantities (see Sections~\ref{sec:results:m11} and~\ref{sec:results:m12}), one can observe some consistent trends on the scaling of the error with the number of trajectories. Firstly, the linear behaviour with a slope of -1/2 on a log-log scale indicates that the error decreases as $1/\sqrt{N_{traj}}$ for sufficiently large $N_{traj}$, as expected for independent trajectories, for all methods. Secondly, for NEMD a smaller error is obtained with a larger applied force ($\sigma_{kl}$ scales as the inverse of the applied force), even though one should use sufficiently small forces to remain in the linear response regime. Then, for both the GK and EH methods, the estimators involving the force introduced in the present work result in a smaller uncertainty compared to standard binning. Note that the benefit of such a force-based estimator increases with decreasing bin width $\Delta z$ (not shown), as discussed in more detail for structural properties\cite{borgisComputationPairDistribution2013,colesComputingThreedimensionalDensities2019}. Finally, the EH approach has a smaller uncertainty compared to GK.

Based on the above findings for both the scaling with the number of trajectories and the computational time per trajectory, we can conclude that the best equilibrium method to compute mobility profiles is the Einstein-Helfand approach, combined with the force-based estimator. Even this method, however, remains less efficient than NEMD for a good choice of applied perturbation, in order to determine a single mobility profile $\mathcal{M}_{kl}$. One should keep in mind, however, that the NEMD approach requires considering several perturbations to find a compromise between a small variance (requiring a large perturbation) and remaining in the linear response regime (\emph{i.e.} sufficiently small perturbation). In addition, several perturbations must be applied to determine all the mobility profiles (each perturbation only provides a row of the mobility matrix), while the equilibrium routes provides them simultaneously. While NEMD seems to be more competitive for the present case of a 2x2 matrix for a binary mixture, the balance should be different for more complex systems. This might already be true for the important case of electrolyte solutions, with at least water, cations and anions and a 3x3 matrix for the responses to pressure, salt concentration and electric potential gradients\cite{yoshida_generic_2014}.

\subsection{Mobility matrix}
\label{sec:results:mall}

\begin{figure}[ht!]
    \includegraphics[width=0.6\columnwidth]{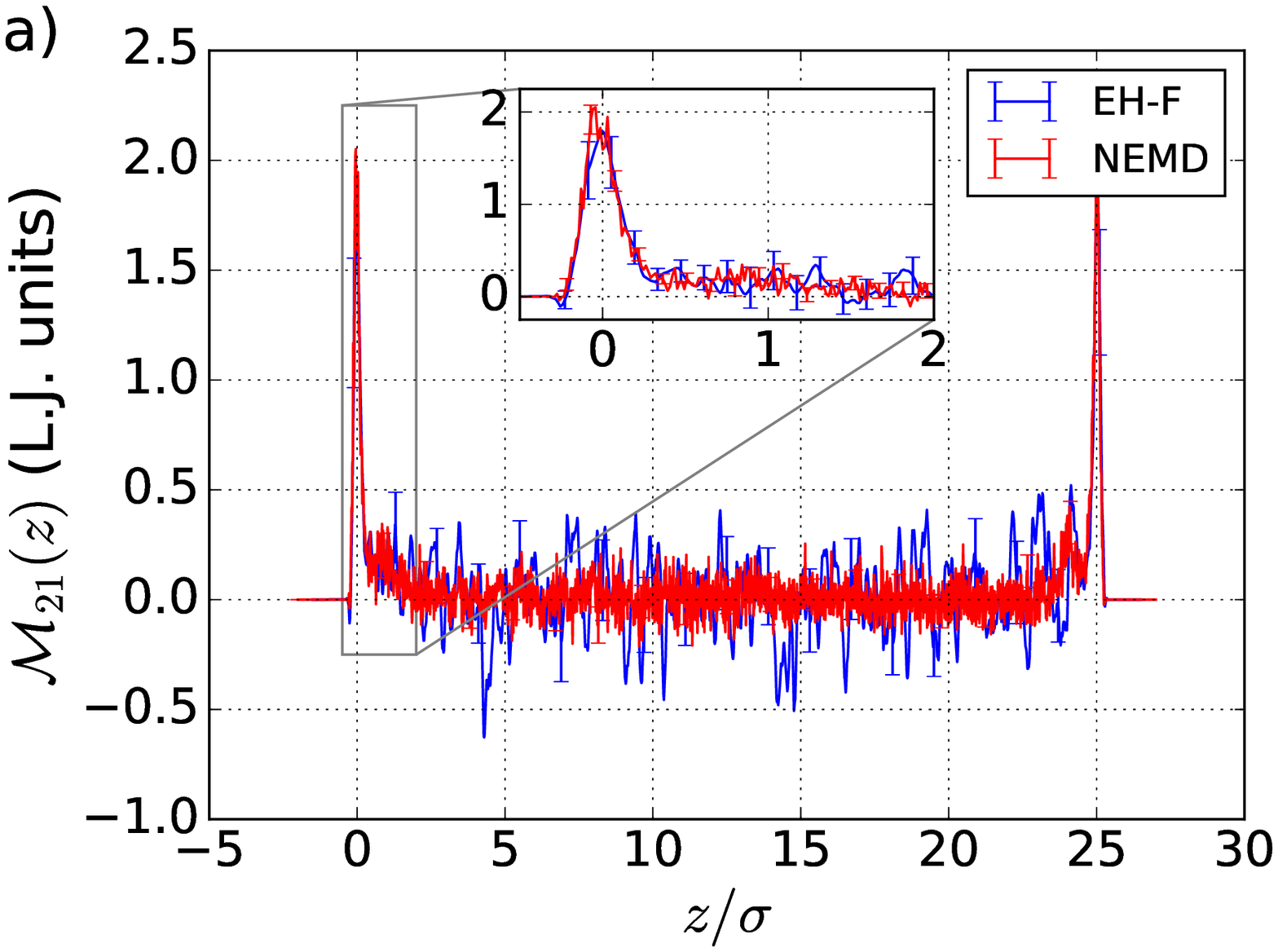}\\
    \includegraphics[width=0.6\columnwidth]{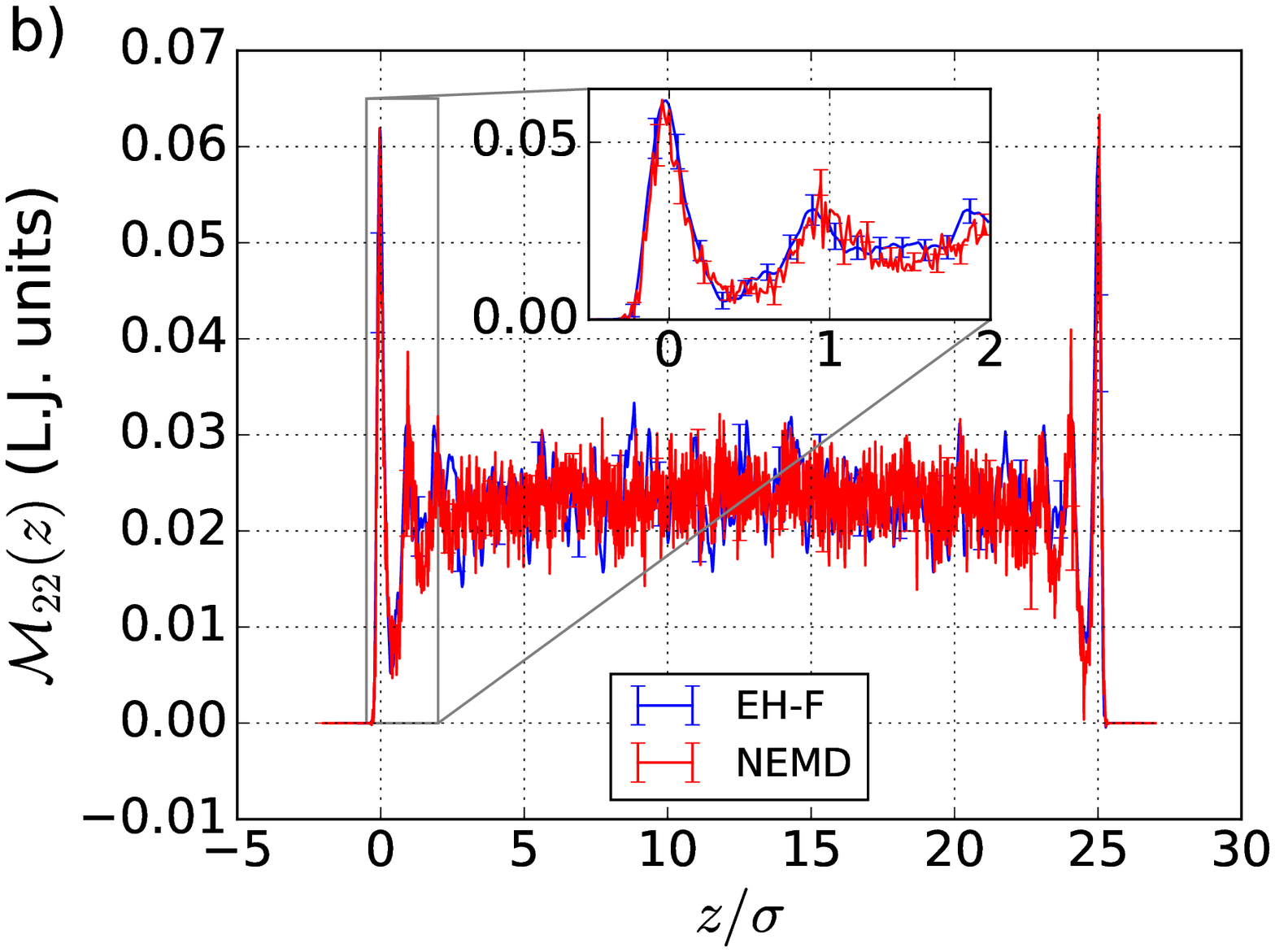}\\
\caption{
Mobility profiles for the excess solute flux in response to (a) a pressure gradient, $\mathcal{M}_{21}$, and (b) a chemical potential gradient, $\mathcal{M}_{22}$. 
For each profile, we show the results obtained by the Einstein-Helfand approach using the mixed estimator involving the forces described in section~\ref{sec:forces:tcf} (blue lines); results of non-equilibrium MD (NEMD) for an external force $f^p=10^{-3}$ and $f^\mu=10^{-2}$ (both in L.J. units) are also indicated as a reference (red lines).
The insets show the same data in the range $[-0.5,2]$.
All results are shown for the same number of independent trajectories ($N_{traj}=150$).}

\label{fig:results:M21M22}
\end{figure}

Figure~\ref{fig:results:M21M22} shows the remaining two mobility profiles, corresponding to the excess solute flux in response to a pressure gradient, $\mathcal{M}_{21}(z)$ and to a chemical potential gradient, $\mathcal{M}_{22}(z)$. As for the diffusio-osmotic response $\mathcal{M}_{12}(z)$, both profiles are flat in the bulk region of the fluid even in the case of the parabolic pressure-induced flow profile (see $\mathcal{M}_{11}(z)$ in Figure~\ref{fig:results:M11}c). This confirms in particular that the mobility profile matrix $\mathcal{M}$ is not symmetric, unlike the Onsager matrix $\mathcal{L}$ for the total fluxes, discussed below. 

Even though it is known that the symmetry of a response matrix depends on the choice of conjugate variables even for global fluxes\cite{degrootNonEquilibrium1984}, we show here using molecular simulations that even when the symmetry is satisfied for the global response, this is not necessarily the case for the local one. The difference between the mobility profiles $\mathcal{M}_{21}(z)$ and $\mathcal{M}_{12}(z)$ can be understood by comparing the definitions of $C_{j_AQ}(t,z)$ and $C_{qJ_A}(t,z)$ in Eqs.~\ref{eq:c21} and~\ref{eq:c12}:
\begin{align}
C_{j_AQ}(t,z) &=  \frac{1}{NS} \langle \sum \limits_{i=1}^{N_A} \sum \limits_{j=1}^{N} v_{x,i}(t)  v_{x,j}(0) \delta (z-z_i(t)) \rangle
\\
&\neq \notag
\\
C_{qJ_A} (t,z) &= \frac{1}{NS} \langle \sum \limits_{i=1}^{N} \sum \limits_{j=1}^{N_A} v_{x,i}(t)  v_{x,j}(0) \delta (z-z_i(t)) \rangle 
\end{align}
From these expressions, one can also note that their integrals over $z\in[0,H]$ are equal, so that $\mathcal{L}_{21}=\mathcal{L}_{12}$ in the Onsager matrix.

We also note that the plateau value in the bulk vanishes for $\mathcal{M}_{21}(z)$ but not for $\mathcal{M}_{22}(z)$. This can be understood by noting that the excess solute flux in the bulk region, where the solute and solvent densities are uniform, can be rewritten using the definitions Eqs.~\ref{eq:q_z}, \ref{eq:j_z} and~\ref{eq:cA} as
\begin{align}
\left\langle j_A^b(z) - c_A^*q^b(z)\right\rangle
& = \frac{\rho_A^b\rho_B^b}{\rho_A^b+\rho_B^b}
\left( \left\langle v_A^b(z)\right\rangle - \left\langle v_B^b(z)\right\rangle \right)
\label{eq:excessfluxbulk}
\end{align}
where the brakets denote a time average, \textit{i.e.} for a non-equilibrium steady-state, and the superscript $b$ refers to observables considered in the bulk region. The excess flux therefore depends on the relative average velocities of the solute and solvent. In the present case where $A$ and $B$ particles are identical (they only differ in their interactions with the wall), the relative velocity vanishes for a pressure gradient (where the perturbation acts identically on both species) but not for a chemical potential gradient (where the perturbation acts in opposite directions for solute and solvent).

\begin{table}[ht!]
\begin{center}
\begin{tabular}{|c|c|}
\hline 
Method  & Onsager matrix $\mathcal{L}$ \\
\multicolumn{1}{c|}{}& (L.J. units) \\
\hline 
Non-equilibrium    & 
$\left(\begin{array}{cc}  
75.3 \pm 0.2 & 0.060 \pm 0.008
\\ 
0.062 \pm 0.003 & 0.0226 \pm 0.0001
 \end{array} \right)$  \\
\hline 
Einstein-Helfand bin   & 
$\left(\begin{array}{cc} 
75.4 \pm 1.8 & 0.062 \pm 0.019
\\
0.062 \pm 0.019 & 0.0231 \pm 0.0005
\end{array} \right)$  \\
\hline 
Einstein-Helfand force   & 
$\left(\begin{array}{cc}  
75.6 \pm 1.8 & 0.062 \pm 0.019
\\ 
0.062 \pm 0.019 & 0.0232 \pm 0.0005
\end{array} \right)$  \\
\hline 
\end{tabular}
\end{center}
\caption{
Coefficients of the Onsager matrix for the total fluxes, obtained by averaging the mobility profiles $\mathcal{M}_{kl}(z)$ over the slit pore (see Eqs.~\ref{eq:onsager}, \ref{eq:tr_mat} and~\ref{eq:LijfromMij}), for the NEMD and with the Einstein-Helfand approach with $N_{traj}=150$ trajectories.
All results are in L.J. units.
}
\label{tab:integrals} 
\end{table}

Finally, Table~\ref{tab:integrals} reports the Onsager matrix $\mathcal{L}$ for the total fluxes, obtained by averaging the mobility profiles $\mathcal{M}_{kl}(z)$ over the slit pore (see Eqs.~\ref{eq:onsager}, \ref{eq:tr_mat} and~\ref{eq:LijfromMij}). The averages and standard error are computed from the averages of the profiles for each trajectory. Of course, there is no need to compute the whole profiles to compute the total fluxes, but we consider the integrals as a final test of consistency for the various methods to determine the mobility profiles. In particular, the number of trajectories necessary to converge the coefficients of the Onsager matrix is smaller than that required to converge the mobility profiles, but we report the results of $N_{traj}=150$ to correspond to the results presented in the previous sections. The results for the NEMD and both Einstein-Helfand methods are fully consistent, as expected from the profiles discussed above. A further important observation is that the off-diagonal terms $\mathcal{L}_{12}$ and $\mathcal{L}_{21}$ are equal, as expected, even though the profiles $\mathcal{M}_{12}(z)$ and $\mathcal{M}_{21}(z)$ are different (see Figures~\ref{fig:results:M12}c and~\ref{fig:results:M21M22}a).

\section{Conclusions}

We have shown how to evaluate mobility profiles for the transport of confined fluids in response to a perturbation from equilibrium molecular simulations, illustrated on the particular case of a binary mixture confined between parallel walls, under pressure or chemical potential gradients. Using linear response theory, we derived the relevant Green-Kubo expressions, which involve time correlation functions between local (solvent and solute) fluxes and global ones. Such correlation functions are difficult to sample accurately, especially for a fine spatial sampling of the mobility profile, and we propose to combine two complementary strategies: on the one hand, we improve the spatial sampling by proposing a mixed estimator of the local fluxes involving not only the positions and velocities of the particles, but also the forces acting on them; on the other hand using the Einstein-Helfand approach (slope of the product of integrated fluxes at long times) instead of the Green-Kubo one (integral of the time correlation function).

We have analyzed in detail the volume flux in response to pressure or chemical potential gradients (Poiseuille and diffusio-osmotic flows, respectively) and compared the performance of all equilibrium methods to the more standard non-equilibrium ones. We recover quantitatively the fluid viscosity and diffusio-osmostic mobility from the mobility profiles in the bulk part of the pore. Completing the analysis with the excess solute fluxes under the same perturbations, we find in particular that the mobility profile matrix is not symmetric, unlike the Onsager matrix for the total fluxes. Such an observation is not unexpected, but doesn't seem to be widely appreciated.

All equilibrium and non-equilibrium methods provide the same mobility profiles, but with different statistical uncertainties. The latter all scale as the inverse square root of the number of independent trajectories, but with different prefactors (which also depend on the considered response). Considering also the computational time per trajectory for each method, we can conclude that the best equilibrium method to compute mobility profiles is the Einstein-Helfand approach, combined with the force-based estimator. This remains less efficient than NEMD for a specific mobility profile $\mathcal{M}_{kl}$, but the NEMD approach requires considering several perturbations to find a compromise between a small uncertainty and remaining in the linear response regime. In addition, several perturbations must be applied in NEMD to determine all the mobility profiles, while the equilibrium routes provide them simultaneously. 

While NEMD seems to be more competitive for the present case of a 2x2 matrix for a binary mixture, the balance should be different for more complex systems. This might already be true for the important case of electrolyte solutions, with at least solvent, cations and anions and a 3x3 matrix for the responses to pressure, salt concentration and electric potential gradients\cite{yoshida_generic_2014}. This would allow in particular to investigate aqueous electrolytes in the pores of charged materials such as clay minerals, as well as through carbon or boron nitride nanotubes, or ultra-narrow slit pores\cite{mouterde_molecular_2019}. The benefit of the equilibrium approach to compute all the responses should in principle be even larger if one also considers thermal gradient (for instance, following strategies of Ref.~\cite{proesmans_comparing_2019}) in addition to the ones discussed above, even though dealing with such perturbations in molecular simulations can be challenging.

\section*{Acknowledgments}
We acknowledge financial support from H2020-FETOPEN project NANOPHLOW (grant number 766972) and from the French Agence Nationale de la Recherche (ANR) under Grant No. ANR-17-CE09-0046- 02 (NEPTUNE). We thank Lyd\'eric Bocquet, Sara Bonella, Giovanni Ciccotti, Rodolphe Vuilleumier, Antoine Carof and Samuel Coles for fruitful discussions.

\section*{Data Availability Statement}
The data that support the findings of this study are available from the corresponding author upon reasonable request.

\appendix

\section{Derivation of eq. (\ref{eq:rho_fsampl})}
\label{app:forces}

In this appendix, we provide the derivation of new estimator involving the force for the one-dimensional density introduced in Section~\ref{sec:forces:density}. 
We first rewrite Eq.~\ref{eq:dens} by adding and subtracting to each term of the sum a coarse-graining function $w_N(z_i-z)$, defined in Section~\ref{sec:forces:density}:
\begin{align}
\rho(z) 
&= \frac{1}{S}  \left[  \left\langle \sum_{i=1}^N  w_N(z_i-z) \right\rangle + \left\langle \sum_{i=1}^N  w'_f(z_i-z) \right\rangle \right]
\label{eq:rhoappend}
\end{align}
where we have introduced the function $w_f$ defined by its derivative 
$w'_f(z)=\delta(z)-w_N(z)$. Specifically, we consider the antiderivative of $\delta(z)-w_N(z)$, given by Eq.~\ref{eq:weightforce}, which vanishes for $|z|>\xi$ (this is possible with our choice of $w_N$, see Eq.~\ref{eq:tria}).
For completeness, we also provide the explicit form of $w_f(z)$ for the particular choice of $w_N(z)$ given by Eq.~\ref{eq:tria}:
\begin{align}
w_f(z) =  \left\lbrace
\begin{array}{l}
 -(\xi+z)^2/2\xi^2 \; \textrm{for}\; z\in[- \xi , 0[ \\ 
  (\xi-z)^2/2\xi^2 \; \textrm{for}\; z\in] 0 , \xi] \\ 
 \textrm{0 otherwise}
 \end{array}  
 \right.
\; , 
\end{align}
but the present derivation is not specific to this choice, provided that $w_N$ satisfies the constraints indicated in the main text.
Recalling the definition Eq~\ref{eq:observable} of the ensemble average, one can for each term $i$ separate in the integral over phase space that over the coordinate $z_i$ from all others (coordinates $x_i$ and $y_i$ of particles, positions ${\bf r}^{N-1}_{j\neq i}$ of particles $j\neq i$ and all momenta ${\bf p}^N$) and integrate by parts. For each of these terms, we obtain:
\begin{align}
\int_{-\infty}^{+\infty} {\rm d}z_i \, e^{-\beta \mathcal{H} (\mathbf{r}^N,\mathbf{p}^N)} w'_f(z_i-z) &= \left[ e^{-\beta \mathcal{H} (\mathbf{r}^N,\mathbf{p}^N)} w_f(z_i-z)\right]_{z_i=-\infty}^{z_i=+\infty} 
\nonumber \\
&  \hspace{-2.5cm}  - \int_{-\infty}^{+\infty} {\rm d}z_i \, \beta f_{z,i} e^{-\beta \mathcal{H} (\mathbf{r}^N,\mathbf{p}^N)} w_f(z_i-z)
\end{align}
where $ f_{z,i}=-\frac{\partial \mathcal{H}}{\partial z_i}$ is the $z$ component of the force acting on particle $i$. The first term vanishes from our choice of function $w_f$ (note that even if we consider $\pm\infty$ to integrate over the full phase space, in practice the walls prevent the particles from leaving the region between 0 and $H$). Integration over all the remaining degrees of freedom then leads to:
\begin{align}
\left\langle \sum_{i=1}^N  w'_f(z_i-z) \right\rangle
&= - \beta \left\langle \sum_{i=1}^N f_{z,i} w_f(z_i-z) \right\rangle
\end{align}
Introducing this result in Eq.~\ref{eq:rhoappend} completes the derivation Eq.~\ref{eq:rho_fsampl}.
A similar strategy can be carried out for the equilibrium cross-correlation functions considered in Section~\ref{sec:forces:tcf}, bearing in mind that an additional term arises from the integration by parts described above (see the main text).

\end{document}